\documentclass[12pt]{article}
\usepackage{graphicx}
\usepackage{amsmath}
\usepackage{amssymb}
\usepackage{graphicx}
\usepackage[utf8]{inputenc}
\usepackage{amsmath}
\usepackage{amsfonts}
\usepackage{amssymb}
\usepackage{graphicx}
\usepackage{color}

\usepackage{hyperref}
\hypersetup{
	hidelinks
}

\begin{document}

	\begin{center}
		\Large {\bf  Gravitational Waves \\ Generated by Null Cosmic Strings}
	\end{center}

	\bigskip
	\bigskip
	
	\begin{center}
		D.V. Fursaev$^\dag$, E.A.Davydov$^{\dag\ddag}$, I.G. Pirozhenko$^{\dag\ddag}$, V.A.Tainov$^{\dag\ddag}$
	\end{center}
	
	\bigskip
	\bigskip
	\date{today}

	\begin{center}
		{\dag \it  Bogoliubov Laboratory of Theoretical Physics\\
			Joint Institute for Nuclear Research\\
			141 980, Dubna, Moscow Region, Russia}\\
			
			and\\
			
		\ddag	{\it Dubna State University, 
			Universitetskaya st. 19\\ }
		\medskip
	\end{center}
	\bigskip

\begin{abstract}
Null cosmic strings  are shown to disturb  gravitational fields of massive bodies and create
outgoing gravitational waves (GW).
Perturbations of the metric caused by a straight null string and a point-like massive source are found as solutions to linearized Einstein equations on a flat space-time.  
An analytic approximation for their asymptotic at future null infinity is derived. A space-time created by the source and the string is shown to have asymptotically 
polyhomogeneous form. We  calculate GW flux in such space-times 
and demonstrate that the averaged intensity of the radiation is maximal in the direction of the string motion.  
Opportunities to detect null string generated gravity waves are briefly discussed. 
\end{abstract}

\newpage

\section{Introduction}\label{intr}

Experimental evidences of the stochastic gravitational background are the subject of active studies by the Advanced LIGO and Virgo collaborations \cite{LIGOScientific:2021nrg}, as well as by radio telescope collaborations, such as  NANOgrav \cite{NANOGrav:2023gor}
and others.
This background, which is formed by gravitational waves produced at different epochs from different sources,
may carry an important information about cosmic strings. For example, the cusps of tensile cosmic strings \cite{Vilenkin:2000jqa} 
which can appear as a result of the Kibble mechanism  \cite{Kibble:1976sj} are known to emit strong beams of high-frequency gravitational waves \cite{Damour:2000wa}.  

In this work we study a scenario when gravitational waves are generated by another type of cosmic strings, which are null or tensionless strings. 
The null strings are one-dimensional objects whose points move  along  trajectories of light rays, orthogonally to strings \cite{Schild:1976vq}.   As a result, the null strings exhibit optical properties: they behave as one-dimensional null geodesic congruences \cite {Fursaev:2021xlm}, may develop caustics  \cite{Davydov:2022qil} etc.

The null cosmic strings, like the tensile
cosmic strings \cite{Vilenkin:2000jqa},\cite{Kibble:1976sj}, are hypothetical astrophysical objects which might have been produced in the very early Universe,  at the Planckian epoch \cite{GM1,GM2,Lee:2019wij,Xu:2020nlh}.  Possible astrophysical and cosmological  effects  of null cosmic strings, such as deviations of light rays in the gravitational field of strings, scattering
of strings by massive sources and etc, are similar to those of the tensile strings \cite{Davydov:2022qil}, \cite{Fursaev:2017aap}, \cite{Fursaev:2018spa}.

As has been recently shown \cite{Fursaev:2022ayo}, \cite{Fursaev:2023lxq}, null cosmic strings disturb fields of 
sources with electric charges or magnetic moments
and produce electromagnetic waves. 
In this work we demonstrate an analogous effect in gravity:
perturbations of gravitational  fields of massive sources caused by
null cosmic strings are radiated away in a form of gravitational waves.

Gravitational field of a straight null cosmic string in a locally flat space-time can be described in terms of holonomies around the string world-sheet \cite{Fursaev:2017aap}, \cite{vandeMeent:2012gb}. The holonomy group is the parabolic subgroup of the Lorentz group,
(or so called null rotations) with the group parameter determined by the string energy per unit  length. The string world-sheet $\mathcal{S}$ belongs to a null hypersurface $\mathcal{H}$ which is the string event horizon. Null rotations, which leave invariant $\mathcal{S}$,  induce  Carroll  transformations of $\mathcal{H}$.

Another way to describe the null string space-time is to consider it as
a shockwave geometry.   Such geometries can be viewed as a result of a 'cut and paste' procedure  \cite{Penrose:1972xrn}, when two copies of a space-time are glued along the shockwave front with a planar supertranslation \cite{tHooft:1985NPB} of one copy. In case 
of the strings the shockwave front is the event horizon $\mathcal{H}$ while the supertranslations are elements of the Carroll group \cite{Duval:2014lpa},\cite{Duval:2014uoa}.

We follow \cite{Fursaev:2017aap}, \cite{Fursaev:2022ayo}, \cite{Fursaev:2023lxq}  
and consider metric perturbations caused by the string as solutions to a characteristic problem \cite{Morse:1953}, when
the linearized Einstein equations are solved for initial data on $\mathcal{H}$. 
The equations are taken on a flat space-time. The initial data are
determined  by a variation of the gravitational field of the source on  $\mathcal{H}$ generated by the Carroll transformation.

A technique to study electromagnetic perturbations caused by straight strings has been elaborated in \cite{Fursaev:2022ayo}, \cite{Fursaev:2023lxq}. In the present paper  we apply it to the gravitational perturbations near the future null infinity. The asymptotic of angular components of the metric at large distances $r$ from the source are shown to have the form:
\begin{equation}\label{i.1}
h_{AB}(r,U,\Omega)\simeq  r \Bigl( C_{AB} (U,\Omega) + \tilde{C}_{AB} (\Omega)\ln (r/\varrho) \Bigr) +O(\ln r)~~,
\end{equation}
where $U$ is a retarded time,  $\Omega$ are spherical coordinates, $\varrho$ is a dimensional parameter related to the approximation.   Symmetric traceless tensors $C_{AB}$, $\tilde{C}_{AB}$ are set in the tangent space of a unit sphere $S^2$,  $C_{AB}$  depends on two polarizations of the gravity wave.
Eq.~\eqref{i.1} holds if $r$ is large enough with respect to $U$ and with respect to an impact parameter between the string and the source.

The  logarithmic term in \eqref{i.1}  appears since standard radiation conditions are violated in the presence of null strings.  Due to this term the resulting geometry, sourced by the string and a point-like mass, belongs
to a class of the so called polyhomogeneous space-times \cite{Chrusciel:1993hx}.  The key property of this class
is in non-analytical behavior at conformal null infinities and in the corresponding modification of the peeling properties of  
the curvature invariants \cite{Friedrich:2017cjg}. It has been realized in last decades, however, that
the polyhomogeneous space-times are important to adopt realistic radiation conditions in General Relativity, see e.g. \cite{Kroon:1998dv},\cite{Capone:2021ouo}.

We show that \eqref{i.1} and the Einstein equations imply a finite total emitted energy of gravity waves
\begin{equation}\label{i.2}
E=\frac{1}{8\pi G} \int^{\infty}_{-\infty} dU\int d\Omega ~N_{AB}  N^{AB} ~~,
\end{equation}
where $N_{AB}$ is an analogue of the Bondi news tensor, $N_{AB}=\frac{1}{2}  \partial_U C_{AB}$, and $A,B$ are risen  with the help of the metric on $S^2$.   The logarithmic terms in \eqref{i.1} do not contribute to $E$. In this work we analytically calculate $N_{AB}$, and 
an angular distribution of the averaged intensity of the radiation.  

The paper is organized as follows.   Formulation of the problem based on the method
developed in  \cite{Fursaev:2017aap}, \cite{Fursaev:2022ayo} is presented in  Sec.~\ref{method}. A short review of null symmetries which allow one to describe holonomy of the null string space-time, as well as the discussion of the related Carroll symmetries of the string event horizon are given in Sec.~\ref{sym}.  
The characteristic problem for perturbations generated by a null string is formulated in Sections~\ref{problem}
and \ref{aux}.  A null string moving in a locally flat space-time does not produce a shock wave. In Section~\ref{Israel} we show that this statement is true (at least in the leading order in perturbations) when the string moves near gravitating sources.   After necessary
preliminaries in Sec.~\ref{prel}, the perturbations caused by a straight string 
and a point-like source without spin are described in Sec.~\ref{elrad}. Their mathematical aspects and
interpretation as outgoing gravity waves are discussed in Sec.~\ref{flux}.  
As is mentioned above, the perturbations have a logarithmic non-analyticity near the future null infinity
and belong to a class of asymptotically polyhomogeneous space-times, see Sec.~\ref{poly}. Properties of the dynamical 
part of metric perturbations, which allow one to identify them with spherical gravity waves, are discussed in Sec.~\ref{strain}. In  Section~\ref{set}  we go beyond linearized approximation to define the stress-energy pseudo-tensor of the gravitational field.
We show that the total radiated energy is finite.
An averaged (over the total observation time) intensity of the radiation is considered in Sec.~\ref{inten}. Like in case of the string generated electro-magnetic radiation
\cite{Fursaev:2022ayo}   the maximum of the intensity of gravity waves is near the line which starts from the source and goes toward  the direction of the string motion.
Since null strings are not straight even in weak gravitational fields we describe,  in Sec.~\ref{rob}, under which conditions 
our results  for straight strings are robust.
We then calculate, in Section~\ref{exp}, the average luminosity to compare it with luminosities of astrophysical objects and to show that effects related to spin (rotation)
of the source are suppressed. 
Perspectives to detect string generated gravity waves, including gravity waves from clumps of dark matter, are discussed in Sec.~\ref{SGWB}.
We finish with a short summary in Sec.~\ref{sum}.
Some parts of our computations, which are lengthy and tedious, are left for Appendix~\ref{App1}.   In Appendix~\ref{app2} 
we demonstrate that averaged $t^r_U$ component of the stress-energy pseudo-tensor is invariant under coordinate transformations  which preserve the leading asymptotic of the metric at null infinities.

\section{Formulation of the problem}\label{method}
\subsection{Strings, shockwaves, and Carroll transformations}\label{sym}
\setcounter{equation}0

In this Section we give necessary definitions by following  notations of \cite{Fursaev:2023lxq}. 
Consider a straight cosmic string which is stretched along $z$-axis and moves along $x$-axis.
The space-time is locally Minkowski, $R^{1,3}$,  with the metric
\begin{equation}\label{1.2}
ds^2=-dv du +dy^2+dz^2~~,
\end{equation}
where $v=t+x$, $u=t-x$. The equations of the string world-sheet $\mathcal{S}$ are  $u=y=0$.
The null rotations $x^\mu=M^\mu_{~\nu}(\lambda)\bar{x}^\nu$ or
\begin{equation}\label{1.3}
u=\bar{u}~~,~~
v=\bar{v}+2\lambda \bar{y}+\lambda^2 \bar{u}~~,~~
y=\bar{y}+\lambda \bar{u}~~,~~z=\bar{z}~~,
\end{equation}
where $\lambda$ is some real parameter, leave invariant \eqref{1.2} and make a subgroup  of the Lorentz group. 
Transformations of quantities with lower indices are
\begin{equation}\label{1.4}
V_u=\bar{V}_u-\lambda \bar{V}_y+\lambda^2 \bar{V}_v~~,~~
V_v=\bar{V}_v~~,~~
V_y=\bar{V}_y-2\lambda \bar{V}_v~~,~~V_z=\bar{V}_z~~,
\end{equation}
or  $V_\mu=M_{\mu}^{\phantom{\mu} \nu}(\lambda )\bar{V}_\nu$, where $M_{\mu}^{\phantom{\mu} \nu}=\eta_{\mu\mu'}\eta^{\nu\nu'}M^{\mu'}_{\phantom{\mu'}\nu'}$.

One can show \cite{Fursaev:2017aap},\cite{vandeMeent:2012gb} that a parallel transport of a vector $V$ along a closed contour around the string
with energy $E$ results in the null rotation $V'=M(\omega)V$, where parameter $\omega$ is defined as
\begin{equation}\label{1.5}
\omega \equiv 8\pi GE~~.
\end{equation}
The world-sheet is a fixed point set of \eqref{1.3}. To avoid confusions let us emphasize that (\ref{1.2}) is not the metric of the string spacetime 
(as compared to (\ref{1.10}), see below). We use (\ref{1.2}) only to set the stage for further analysis.

The null hypersurface $u=0$ is the event horizon of the string $\mathcal{H}$. We enumerate coordinates $v,y,z$ on $\mathcal{H}$
by indices $a,b,\ldots$ 
Coordinate transformations \eqref{1.3}  at $u=0$ (with $\lambda=\omega$) induce the change  of coordinates  on $\mathcal{H}$:
\begin{equation}\label{1.14a}
x^a=C^{a}_{~b}\bar{x}^b~~~\mbox{or}~~~{\bf x}=\bar{\bf x}+2\omega y{\bf{q}}~~,~~q^a=\delta^a_v~~,
\end{equation}
where ${\bf x}\equiv \{v,y,z\}$.  Matrix  $C^{a}_{~b}$ can be defined as
\begin{equation}\label{1.14b}
C^{a}_{~b}=M^{a}_{~b}(\omega)~~.
\end{equation}
Transformations \eqref{1.14a}, \eqref{1.14b}  make the Carroll group \cite{Duval:2014lpa},\cite{Duval:2014uoa}. A review of Carroll transformations
and their applications can be found in \cite{Ciambelli:2023tzb},\cite{Ciambelli:2023xqk}.
The components of one-forms $\theta=\theta_a dx^a$ and vector fields $V=V^a\partial_a$ on $\mathcal{H}$, in a coordinate  
basis, change as
\begin{equation}\label{1.18a}
\theta_a({\bf x})=C_{a}^{~b}~\bar{\theta}_b(\bar{\bf x})~~,~~V^a({\bf x})=C^{a}_{~b}~\bar{V}^b(\bar{\bf x})~~,
\end{equation} 
where $C_{a}^{~b}=M_{a}^{~b}$.
Since $M_b^{~u}=0$ the matrix  $C_{a}^{~b}$ is inverse of $C^{a}_{~c}$. 
It should be noted, however, that indices $a,b$ cannot be risen or lowered since the metric of $\mathcal{H}$ is degenerate.

By following the method suggested in \cite{Fursaev:2017aap} one can construct a string space-time, which is locally flat but has the required holonomy  on $\mathcal{S}$.
One starts with $R^{1,3}$ which 
 is decomposed onto two parts: $u<0$ and $u>0$.  Trajectories of particles and light rays at $u<0$ and $u>0$
can be called ingoing and outgoing trajectories, respectively.
To describe outgoing trajectories, one introduces two types
of coordinate charts: $R$- and $L$-charts, with cuts on the horizon either on the left
($u=0, y<0$) or on the right ($u=0, y>0$). The initial data on the string horizon are related to the ingoing data by the 
Carroll transformations \eqref{1.14a}. For  brevity the right  ($u=0, y>0$) and the left ($u=0, y<0$) parts of $\mathcal{H}$ will be denoted as 
$\mathcal{H}_+$
and $\mathcal{H}_-$.

For the $R$-charts the cut is along $\mathcal{H}_-$. If $x^\mu$ and $\bar{x}^\mu$ are, respectively, the coordinates above and
below the horizon the transition conditions on $\mathcal{H}$ in the $R$-charts look as
\begin{equation}\label{1.6}
x^a\mid_{\mathcal{H}_+}=\bar{x}^a~~,
\end{equation}
\begin{equation}\label{1.7}
x^a\mid_{\mathcal{H}_-}=C^a_{~b}\bar{x}^b~~.
\end{equation}
Analogously, 3 components of 4-velocities of particles and light rays (with the lower indices)
change on $\mathcal{H}_-$ as $u_a=C_a^{~b}\bar{u}_b$. 

Coordinate transformations \eqref{1.7} are reduced to a linear supertranslation of the single coordinate
\begin{equation}\label{1.8}
v=\bar{v}+2\omega y~~,~~y<0~~.
\end{equation}

As a result of the Lorenz invariance of the theory the descriptions based on $R$- or $L$-charts are equivalent. 
Without loss of the generality from now on we work with the $R$-charts. 
In this case all field variables experience Carroll transformations on  $\mathcal{H}_-$,
and one needs to solve field equations in the domain $u>0$ with the initial data changed on $\mathcal{H}_-$.

The gravitational field of a null string can be also described by the metric
\begin{equation}\label{1.10}
	ds^2=-dv du - \omega |y| \delta(u)du^2+dy^2+dz^2~~,
\end{equation} 
where $\omega$ is defined by \eqref{1.5}. This string space-time is locally flat, except the world-sheet, where 
the $uu$ component of the Ricci tensor  has a delta-function singularity.
The delta-function in \eqref{1.10} reflects the fact that the standard coordinate chart is not smooth at $\mathcal{H}$.  Geometry  \eqref{1.10} is a particular case of gravitational shockwave backgrounds:
\begin{equation}\label{1.11}
	ds^2=-dv du +f(y)\delta(u)du^2+\sum_i dy_i^2~~,~~i=1,...n~~.
\end{equation}
The hypersurface  $u=0$ is the shockwave front.
Shockwaves \eqref{1.11} are exact solutions of the Einstein equations sourced by a stress energy tensor localized at $u=0$ and 
having the only non-vanishing $uu$ component.

Penrose  \cite{Penrose:1972xrn} viewed \eqref{1.10}  as two copies  
of Minkowski space-times glued along the hypersurface $u=0$ with the
shift of the $v$-coordinate of the upper copy ($u>0$) to $v-f(y)$. 
This prescription is equivalent to approach discussed above.
Consider a coordinate chart with the cut along the entire surface $\mathcal{H}$ and choose the Penrose coordinate transformations:
\begin{equation}\label{1.12}
	v=\bar{v}+\omega |y|~~,~~u=0~~.
\end{equation} 
By comparing \eqref{1.12} with \eqref{1.8} one can see that coordinates on  $\mathcal{H}_-$ are Carroll-rotated by the 'angle' $\omega/2$
and by $-\omega/2$ on $\mathcal{H}_+$.  That is, the relative transformation on $\mathcal{H}_\pm$ is the rotation by 'angle' $\omega$,
as it should be. 

Penrose transformations  for shockwaves are called "planar supertranslations". This should not be confused with supertranslations of the BMS group. In case of null strings \eqref{1.12} belongs to the Lorentz subgroup of the BMS group.

Instead of the approach we use in the present paper one may try to study perturbations caused by a massive source by using (\ref{1.10}) as a background metric of the string spacetime. This method would meet serious technical challenges because of the delta-function singularities. A mathematically rigorous approach would be to introduce a regularization and show that a final result is finite and regularization independent when the regularization is removed.
This seems to be a more complicated task.

\subsection{Characteristic problem}\label{problem}

Consider a space-time ${\mathcal{M}}_{\mbox{\tiny{in}}}$ whose metric is a solution to linearized Einstein equations
in the absence of the string
\begin{equation}\label{1.13}
g^{\mbox{\tiny{in}}}_{\mu\nu}\simeq \eta_{\mu\nu}+\hat{h}_{\mu\nu}~~.
\end{equation} 
Here $\hat{h}_{\mu\nu}$ is a first correction to the flat metric $\eta_{\mu\nu}$ due to the presence of a point mass.
We use (\ref{1.13}) to describe the metric below the string horizon.

A null cosmic string causes a perturbation $h_{\mu\nu}$ so that the new metric above the string horizon (in the future of
$\mathcal{H}$) takes the form:
\begin{equation}\label{1.14}
g^{\mbox{\tiny{out}}}_{\mu\nu}=g^{\mbox\tiny{in}}_{\mu\nu}+h_{\mu\nu}~~.
\end{equation} 
We denote the corresponding manifold by $\mathcal{M}_{\mbox{\tiny{out}}}$.  In the linearized approximation
\begin{equation}\label{3.1}
g^{\mbox{\tiny{out}}}_{\mu\nu}\simeq \eta_{\mu\nu}+h^{(1)}_{\mu\nu}~~,~~h^{(1)}_{\mu\nu}=\hat{h}_{\mu\nu}+h_{\mu\nu}~~.
\end{equation} 
If the trajectory of a point mass is not affected by the string
(in a suitable coordinate chart, see below)
then outside the string world-sheet $g^{\mbox{\tiny{out}}}$ and
$g^{\mbox{\tiny{in}}}$ are solutions of linearized Einstein equations with the same source. Thus, problems for $\hat{h}_{\mu\nu}$ at $u<0$ and 
for $h^{(1)}_{\mu\nu}$ at $u<0$  can be written as
\begin{equation}\label{1.16}
-\Box \left(\hat{h}_{\mu\nu}-\frac 12 \eta_{\mu\nu}\hat{h}\right)=\kappa T_{\mu\nu}~~,~~
\partial^\mu\left(\hat{h}_{\mu\nu}-\frac 12 \eta_{\mu\nu}\hat{h}\right)=0~~,
\end{equation} 
\begin{equation}\label{1.16b}
-\Box \left(h^{(1)}_{\mu\nu}-\frac 12 \eta_{\mu\nu}h^{(1)}\right)=\kappa T_{\mu\nu}~~,~~
\partial^\mu\left(h^{(1)}_{\mu\nu}-\frac 12 \eta_{\mu\nu}h^{(1)}\right)=0~~,
\end{equation} 
with $\Box=-4\partial_u\partial_v+\partial_y^2+\partial_z^2$. Here $\hat{h}=\eta^{\mu\nu}h_{\mu\nu}$, $h^{(1)}=\eta^{\mu\nu}h^{(1)}_{\mu\nu}$, $\kappa=16\pi G$. The stress-energy tensor
$T_{\mu\nu}$ has a delta-function-like form with a support on the trajectory of the source. Here and in what follows we use the harmonic gauge since it is Lorentz invariant.

If we use decomposition $h^{(1)}_{\mu\nu}=\hat{h}_{\mu\nu}+h_{\mu\nu}$ and  Eq. (\ref{1.16}) the string caused perturbations $h_{\mu\nu}$ satisfy the homogeneous linearized equations:
\begin{equation}\label{1.20a}
\Box h_{\mu\nu}=0~~,~~\partial^\mu\left(h_{\mu\nu}-\frac{1}{2} \eta_{\mu\nu} h\right)=0~~,
\end{equation} 
To fix the solution of \eqref{1.20a} one needs initial data on $\mathcal{H}$.

In coordinates \eqref{1.2}  the string world-sheet $\mathcal{S}$ in the leading approximation will be taken in the previous form,
as $u=y=0$, and equation of $\mathcal{H}$ as $u=0$. It should be noted that  $u=0$ is not null for the perturbed metrics
unless $\hat{h}^{uu}=h^{uu}=0$. Although even weak perturbations may cause drastic deformations 
of the string trajectory (for example, they may create caustics) \cite{Davydov:2022qil} we ignore them in our approximation. 
Further comments on the role of deformations of the string are given in Section \ref{rob}.

A new geometry sourced by the point mass and the string is obtained by soldering along $\mathcal{H}$ the part of $\mathcal{M}_{\mbox{\tiny{in}}}$, which is in the past of $\mathcal{H}$, and $\mathcal{M}_{\mbox{\tiny{out}}}$, which is in the future of  $\mathcal{H}$. As has been discussed in Section \ref{sym},
the soldering must be accompanied by the Carroll transformation of $g^{\mbox{\tiny{in}}}$. That is, in the $R$-chart the condition 
looks as 
\begin{equation}\label{1.21a}
g^{\mbox{\tiny{out}}}_{ab}(u=0, {\bf x})=g^{\mbox{\tiny{in}}}_{ab}(u=0, {\bf x})~~,~~y>0~~,
\end{equation}
\begin{equation}\label{1.21b}
g^{\mbox{\tiny{out}}}_{ab}(u=0, {\bf x})=C_{a}^{~c}~C_{b}^{~d} ~g^{\mbox{\tiny{in}}}_{cd}(u=0, \bar{\bf x})~~,~~y<0~~,
\end{equation} 
\begin{equation}\label{1.23}
{\bf x}=\{v,y,z\}~~,~~\bar{\bf x}={\bf x}-2\omega y{\bf{q}}~~,~~q^a=\delta^a_v~~.
\end{equation}
where $C_{a}^{~c}$ are inverse Carroll matrix \eqref{1.14b}.  By taking into account \eqref{1.14} and the fact that
Carroll  transformations are isometries of $\mathcal{H}$, these conditions imply
the following initial data for string inspired perturbations:
\begin{equation}\label{1.21c}
h_{ab}(u=0, {\bf x})=0~~,~~y>0~~,
\end{equation}
\begin{equation}\label{1.21d}
h_{ab}(u=0, {\bf x})=C_{a}^{~c}~C_{b}^{~d}~ \hat{h}_{cd}(\bar{\bf x})-\hat{h}_{ab}({\bf x})~~,~~y<0
~~,
\end{equation} 
where $\hat{h}_{ab}({\bf x})$ are the values of $\hat{h}_{ab}(x)$ on $\mathcal{H}$. At small $\omega$ conditions \eqref{1.21d} are reduced to 
the Lie derivative, $h_{ab}(u=0, {\bf x})\simeq \omega \mathcal{L}_\chi \hat{h}_{ab}({\bf x})$, generated by the vector field $\chi^a=-2yq^a$.

If one prefers to work in the $L$-chart the metric of $\mathcal{M}_{\mbox{\tiny{out}}}$ should be changed to 
$$
\tilde{g}^{\mbox{\tiny{out}}}_{\mu\nu}(x)=M_{\mu}^{\phantom{\mu}\lambda}(-\omega)M_{\nu}^{\phantom{\nu}\rho}(-\omega)
g^{\mbox{\tiny{out}}}_{\lambda\rho}(x)~~.
$$
In this chart the soldering conditions on $\mathcal{H}$ are dual to \eqref{1.21a}, \eqref{1.21b}. Now components with indices $a,b$ are continuous on $\mathcal{H}_-$ but undergo Carroll transformations on $\mathcal{H}_+$.

In this work we consider a point-like source with mass $M$. 
We assume that the source is at rest at a point with coordinates $x_o=z_o=0,y_o=a>0$. Since the string trajectory is $x=t$, $a$ is an impact parameter between the string and the source. We take
\begin{equation}\label{1.17}
\hat{h}_{00}=r_g \phi~~~,~~~\hat{h}_{ij}=r_g \delta_{ij} \phi~~,~~\hat{h}_{0i}=0~~,
\end{equation}
where $i,j,k$ correspond $x,y,z$, $r_g=MG$, and
\begin{equation}\label{1.18}
\phi(x,y,z)=\frac{1}{\sqrt{x^2+(y-a)^2+z^2}}~~.
\end{equation}
One can check that \eqref{1.17} satisfy (\ref{1.16}) and derive explicit form of $T_{\mu\nu}$.
In Section~\ref{exp} we also discuss effects when \eqref{1.17} includes components related  to the spin of the source.

\subsection{Auxiliary problem}\label{aux}

For further purposes it is convenient to introduce components
\begin{equation}\label{1.24}
\bar{h}_{\mu\nu}=h_{\mu\nu}-\frac 12 \eta_{\mu\nu} h~~,
\end{equation}
\begin{equation}\label{1.25}
\Box \bar{h}_{\mu\nu}=0~~,~~\partial^\mu \bar{h}_{\mu\nu}=0~~.
\end{equation} 
Accordingly, we redefine perturbation caused by the mass
\begin{equation}\label{1.26}
\hat{\bar{h}}_{\mu\nu}=\hat{h}_{\mu\nu}-\frac 12 \eta_{\mu\nu} \hat{h}~~,~~
\hat{\bar{h}}_{00}=2r_g \phi~~,~~\hat{\bar{h}}_{ij}=\hat{\bar{h}}_{0i}=0~~,
\end{equation}
where indices $i,j,k$ correspond to $x,y,z$.
Consider the following auxiliary characteristic problem:
\begin{equation}\label{1.27}
\Box \bar{h}_{ab}=0~~,
\end{equation} 
\begin{equation}\label{1.28}
\bar{h}_{ab}(u=0, {\bf x})=0~~,~~y>0~~,
\end{equation} 
\begin{equation}\label{1.29}
\bar{h}_{ab}(u=0, {\bf x})=C_{a}^{~c}~C_{b}^{~d}~ \hat{\bar{h}}_{cd}(\bar{\bf x})-\hat{\bar{h}}_{ab}({\bf x})~~y<0
~~.
\end{equation} 
If solution of \eqref{1.27}-\eqref{1.29} is known the rest components of $\bar{h}_{u\mu}$ can be obtained from the 
gauge conditions $\partial^\mu \bar{h}_{\mu\nu}=0$.  

Since $\Box=-4\partial_u\partial_v+....$ the derivatives of fields over $u$ at $u=0$ are not independent, and the initial data include only values of fields. Thus, we come to the characteristic problem for 6 components $\bar{h}_{ab}$.
In the harmonic gauge there are 4 residual coordinate transformations generated
by a vector field $\chi^\mu$ provided that $\Box \chi^\mu=0$. Thus, the characteristic initial value problem depends on 2 physical degrees of freedom.

Let us show
that the solution $\bar{h}_{\mu\nu}$ of the auxiliary problem
results, with the help of \eqref{1.24}, in the solution $h_{\mu\nu}$  of our problem with conditions \eqref{1.21c}, \eqref{1.21d}.
To do this we note that the solution of the auxiliary problem can be written as
\begin{equation}\label{1.30}
\bar{h}_{\mu\nu}(x)=M_{\mu}^{\phantom{\mu}\lambda}(\omega)M_{\nu}^{\phantom{\nu}\rho}(\omega)\Phi_{\lambda\rho}(\bar{x})-\Phi_{\mu\nu}(x)~~,
\end{equation} 
where $\bar x$ are defined by \eqref{1.3}, and
\begin{equation}\label{1.31}
\Box \Phi_{\mu\nu}=0~~,~~\partial^\mu \Phi_{\mu\nu}=0~~,
\end{equation} 
\begin{equation}\label{1.32}
\Phi_{ab}(u=0, {\bf x})=0~~,~~y>0~~,
\end{equation} 
\begin{equation}\label{1.33}
\Phi_{ab}(u=0, {\bf x})=\hat{\bar{h}}_{ab}({\bf x})~~,~~y<0~~.
\end{equation} 
The proof of \eqref{1.30} follows from the Lorentz invariance of \eqref{1.31}  and relation between null rotations 
$M_{\mu}^{\phantom{\mu}\nu}(\omega)$ and the Carroll transformations $C_{a}^{~b}$.

The values of $\Phi_{u\mu}$ on $\mathcal{H}_-$ can be found from the gauge conditions and equations (\ref{1.31}), 
\begin{equation}\label{1.34}
\Phi_{ua}(0,v,y,z)=\frac 12\int^{v}_{-\infty}\biggl(
\partial_i\Phi_{ia}(0,v',y,z)- \frac 12 \int^{v'}_{-\infty}\partial_i^2 \Phi_{va}(0,v'',y,z)dv''\biggr)dv'~~,
\end{equation} 
\begin{equation}\label{1.35}
\Phi_{uu}(0,v,y,z)=\frac 12\int^{v}_{-\infty} \biggl(\partial_i\Phi_{iu}(0,v',y,z)-2\partial_u \Phi_{vu}(0,v',y,z)\biggr)dv'~~,
\end{equation} 
where $i=y,z$. The right hand side of \eqref{1.35} is determined by \eqref{1.34}.  Integrals in \eqref{1.34}, \eqref{1.35}
converge at $v\to -\infty$ for $y<0$. On the base of \eqref{1.33} one concludes that 
\begin{equation}\label{1.36}
	\begin{split}
		\Phi_{u\mu}(u=0, {\bf x})=0~~,~~y>0~~, \\
		\Phi_{u\mu}(u=0, {\bf x})=\hat{\bar{h}}_{u\mu}({\bf x})~~,~~y<0~~.
	\end{split}
\end{equation}
Another way to get $\Phi_{u\mu}$ on $\mathcal{H}_-$ is to note that the gauge conditions and equations for perturbations 
$\hat{\bar{h}}_{\mu\nu}$ near $\mathcal{H}_-$  coincide with those for $\Phi_{\mu\nu}$.
That is, \eqref{1.33} imply \eqref{1.36}.

Now, by using Eqs.~\eqref{1.24}, \eqref{1.30}, \eqref{1.33} and \eqref{1.36} we prove that our solution does obey
the required conditions \eqref{1.21c}, \eqref{1.21d}. This means that instead of solving the problem formulated in Sec.~\ref{problem}
it is enough to deal with more simple problems  \eqref{1.25}-\eqref{1.29} or  \eqref{1.31}-\eqref{1.33}.

\subsection{Shock-wave geometry without a shock wave}\label{Israel}

Soldering space-times along null hypersurfaces is an interesting technique in general relativity which has been studied in a number of publications to describe null shells and, in particular, shock-wave geometries.
The soldering results in jumps of the components of the Riemann tensor on the null hypersurface.
The jumps determine the Barrabes-Israel stress-energy tensor of the surface layer \cite{Barrabes:1991ng}, which
yields the surface energy density and the pressure of shock waves.
In fact, two given space-times  can be soldered along a null surface by infinitely many physically different ways \cite{Blau:2015nee}. 

In this Section we show that our conditions \eqref{1.26} do not result in jumps of the Riemann tensor and therefore no shock waves
appear whose wave front carry any energy.  We follow \cite{Poisson:2002nv} where the jumps of the curvature 
have been defined in terms of the transverse curvature tensor
\begin{equation}\label{1.37}
\mathcal{C}_{ab}=e^\mu_a e^\nu_b n_{\mu;\nu}~~,
\end{equation} 
where $e_a$ are vectors tangent to the null hypersurface, $e_v=l$ and $n$ are light-like, and $(n\cdot l)= -1$, $(n\cdot e_i)=0$,
see details in \cite{Poisson:2002nv}.
The Barrabes-Israel stress-energy tensor is non-trivial if 
$[\mathcal{C}_{ab}]\equiv \mathcal{C}^{\mbox{\tiny{in}}}_{ab}-\mathcal{C}^{\mbox{\tiny{out}}}_{ab}\neq 0$.
As is easy to see, the surface stress-energy tensor in the considered case is (up to unimportant normalization)
\begin{equation}\label{1.38}
[\mathcal{C}_{ab}]=h_{ab,u}-h_{au,b}-h_{bu,a}~~.
\end{equation} 
It is convenient to use the $R$-chart ($L$-chart) to show that 
$[\mathcal{C}_{ab}]=0$ on $\mathcal{H}_+$ ($\mathcal{H}_-$). 
In the $R$-chart it is enough to show that 
\begin{equation}\label{1.39}
\bar{h}_{ab,u}(u=0,{\bf x})=\bar{h}_{au,b}(u=0,{\bf x})=\bar{h}_{uv,u}(u=0,{\bf x})=0~~,~~y>0~~.
\end{equation} 
Conditions $\bar{h}_{ab,u}=0$ follow from $\Box \bar{h}_{ab}=0$. Given this and gauge equations $\partial^\mu \bar{h}_{\mu a}=0$
one concludes that $\partial_v \bar{h}_{ua}=0$. If we require that $h_{ua}$ vanishes at $|v|\to \infty$, then 
$\bar{h}_{ua}=0$ on $\mathcal{H}_+$.
It follows then from $\Box \bar{h}_{ua}=0$ that $\bar{h}_{ua,u}=0$. Thus $[\mathcal{C}_{ab}]=0$ on $\mathcal{H}_+$.
Analogous arguments can be used in the $L$-chart to show that $[\mathcal{C}_{ab}]=0$ on $\mathcal{H}_-$.

\section{Perturbations}\label{Asympt}
\subsection{Preliminaries}\label{prel}
\setcounter{equation}0

As has been explained in Section~\ref{aux} perturbations caused by the string can be found from solutions 
to characteristic problem \eqref{1.27}-\eqref{1.29}. We show that the perturbations have a form of outgoing  gravitational waves
with asymptotic \eqref{i.1}. To describe this asymptotic behavior it is convenient to go from \eqref{1.2} to retarded time coordinates
\begin{equation}\label{2.1}
	ds^2= -dU^2-2dUdr+r^2d\Omega^2~~, 
\end{equation}
where $U=t-r$ and $r=\sqrt{x^2+y^2+z^2}$. We denote coordinates $\theta, \varphi$ on the unit sphere by $x^A$, $d\Omega^2=\gamma_{AB}dx^Adx^B=\sin^2\theta d\varphi^2+d\theta^2$. 
Our results will be expressed in terms of a unit vector $\vec{n}$ with components $n_x=x/r, n_y=y/r, n_z=z/r$.  If $\vec{l}$ is a unit
vector along the velocity of the string, $\vec{p}$ is another unit vector along the string axis ($l^i=\delta^i_x$, $p^i=\delta^i_z$), then
$n_x=(\vec{n}\cdot \vec{l})$, $n_z=(\vec{n}\cdot \vec{p})$.
We use coordinates \eqref{2.1} above the string 
horizon, $u>0$. In this region $U>-r(1-n_x)$. The region includes the domain $U>0$. At $r\to \infty$ no restrictions on $U$ appear.

\begin{figure}
	\begin{center}
	\includegraphics[width=6cm]{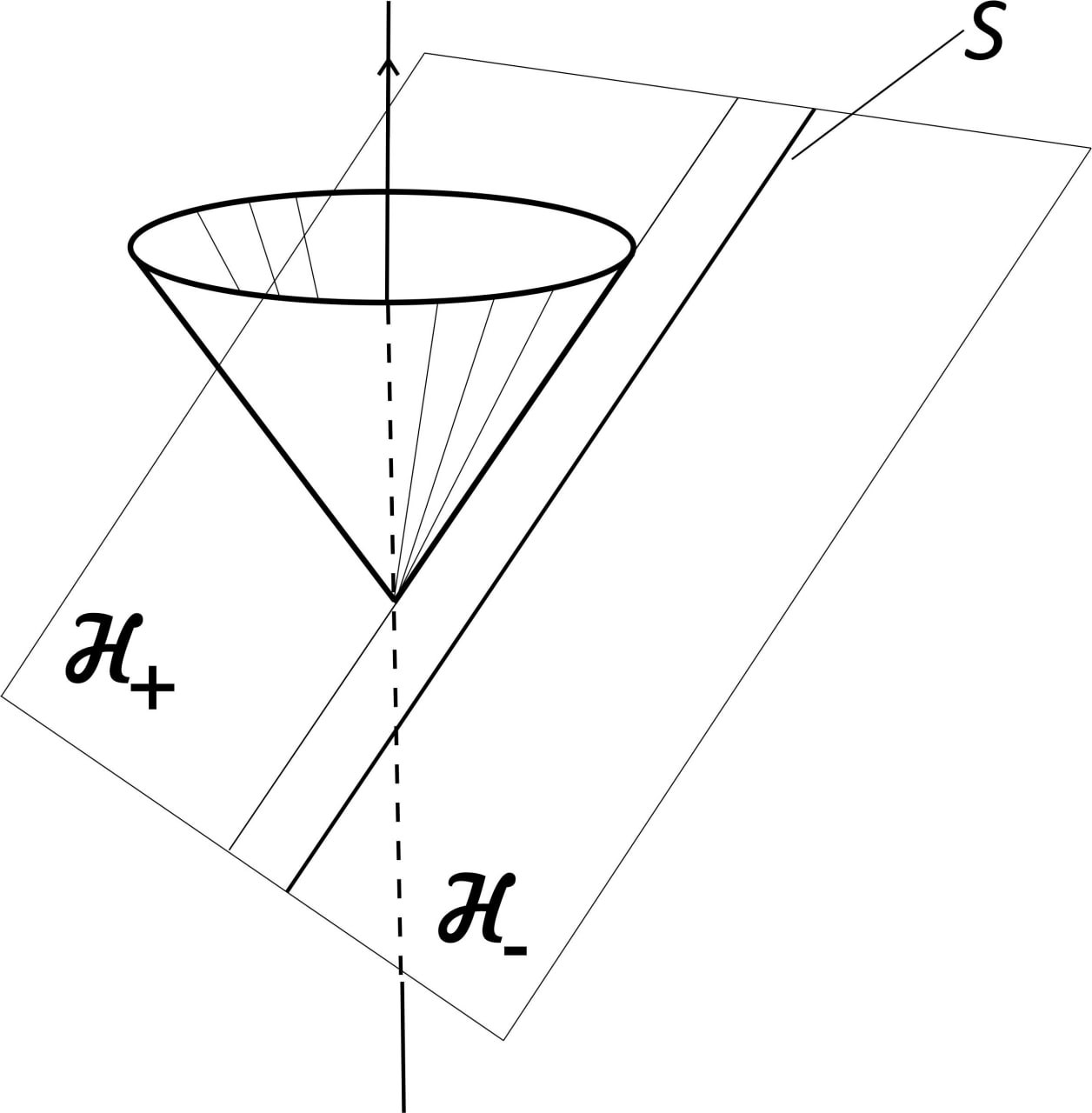}
	\end{center}
	\caption{demonstrates the world-sheet of the string  $\mathcal{S}$, the string horizon $\mathcal{H}=\mathcal{H}_+\cup \mathcal{H}_-$ and
a trajectory of point-like mass (the dashed line) that crosses  $\mathcal{H}_+$.  At late times gravity waves generated
by the string and the mass propagate along null cones $U=C$. Shown on the figure 
is the cone $U=0$ which is tangent to $\mathcal{H}$ and has the apex approximately at a point where the trajectory of the source crosses the horizon.}
\label{Cone}
\end{figure}

The maximum of the luminosity is close to the moment $U=0$. Position of null surfaces $U=0$, $u=0$, the string world-sheet,
and the trajectory of the point mass are shown on Fig.~\ref{Cone}.

As we will see, solutions to problem \eqref{1.27}-\eqref{1.29} can be generated by a solution of the following scalar problem:
\begin{equation}\label{2.2}
	\Box \Phi_\omega(u, {\bf x})=0~~,~~\Phi_\omega(0, {\bf x})=\theta(-y)~ f_\omega({\bf x})~~,
\end{equation} 
\begin{equation}\label{2.3}
f_\omega({\bf x})\equiv f({\bar {\bf x}})~~,~~f({\bf x}) \equiv  \phi(v/2,y,z)~~.
\end{equation}
Here ${\bar {\bf x}}$ is defined in \eqref{1.23}, and $\phi(x,y,z)$ in \eqref{1.18}. The solution has a useful 
integral representation found in  \cite{Fursaev:2022ayo}:
\begin{equation}\label{2.4}
\Phi_\omega(x)=-C \int_{S^2}d\Omega'~\Re\left(
\frac{\tilde{\Phi}_\omega (\Omega')}{x \cdot m (\Omega') +ia\varepsilon(\Omega')}\right)
	~~,~~\tilde{\Phi}_\omega (\Omega')\equiv\frac{\cos\varphi'}{g(\Omega',\omega)}~~.
\end{equation}
where $C=1 /(8\pi^3)$. The integration goes over a unit sphere $S^2$, with coordinates $\Omega'=(\theta',\varphi')$, 
$d\Omega'=\sin\theta' d\theta d\varphi'$. Other notations are:
\begin{align}
	m_u=1-\sin^2\theta'\cos^2\varphi'~~&,~~
	m_v=\sin^2\theta'\cos^2\varphi'~~, \nonumber
	\\
	m_y=\sin 2\theta' \cos\varphi'~~&,~~
	m_z=\sin^2\theta'\sin 2\varphi'~~, 
	\\
	g(\Omega',\omega)=e^{i\theta'}+\omega \sin\theta' \cos\varphi'~~&,~~\varepsilon(\Omega')=2\sin^2\theta'\cos\varphi'~~. \nonumber
\end{align}
One can check that vector field $m_\mu$ is null, $m^2=0$, which guarantees that $\Box \Phi_\omega=0$.

\subsection{Exact form of perturbations and asymptotics}\label{elrad}

Boundary components of the metric of the source,
which determine conditions \eqref{1.29},  are
\begin{equation}\label{2.7}
\hat{h}_{vv}({\bf x})=\frac{r_g}{2} f({\bf x})~~,~~\hat{h}_{vi}({\bf x})=\hat{h}_{ij}({\bf x})=0~~,~~i,j=y,z~~,
\end{equation}
\begin{equation}\label{2.8}
	\begin{split}
		\bar{h}_{vv}(u=0, {\bf x})=\frac{r_g}{2} \theta(-y) \bigl(f_\omega({\bf x})-f({\bf x})\bigr)~~&,~~
		\bar{h}_{yy}\mid_{\mathcal{H}_-}=2r_g  \theta(-y)
		\omega^2 f_\omega({\bf x})~~,
		\\
		\bar{h}_{vy}\mid_{\mathcal{H}_-}=-\omega r_g \theta(-y) f_\omega({\bf x})~~&,~~
		\bar{h}_{za}\mid_{\mathcal{H}_-}=0~~,
	\end{split}
\end{equation}
where  $f({\bf x}), f_\omega({\bf x})$ are defined in \eqref{2.3}. 
Equations~\eqref{2.8} can be obtained with the help of \eqref{1.4}.

Solution $\bar{h}_{ab}(x)$ to \eqref{1.27}-\eqref{1.29} can be expressed by using \eqref{2.2} as
\begin{equation} \label{2.10}
	\begin{split}
		\bar{h}_{vv}(x)=\frac{r_g}{2} \bigl(\Phi_\omega(x)-\Phi(x)\bigr)~~&,~~
		\bar{h}_{yy}(x)=2r_g  \omega^2 \Phi_\omega(x)~~, 
		\\
		\bar{h}_{vy}(x)=-\omega r_g \Phi_\omega(x)~~&,~~
		\bar{h}_{za}(x)=0~~,
	\end{split}
\end{equation}
where $\Phi(x) = \Phi_{\omega=0}(x)$.
With gauge conditions and Eqs.~\eqref{2.10} one can construct other components $\bar{h}_{\mu\nu}$.
As a result the string generated perturbations at $u>0$ can be represented in the following integral form:
\begin{equation}\label{2.12}
h_{\mu\nu} (x)=- C \int_{S^2} d\Omega'~\Re\left({\beta_{\mu\nu} (\Omega')\over x\cdot m(\Omega')+
ia\varepsilon(\Omega') }\right)
~~,
\end{equation}
\begin{equation}\label{2.13}
\beta_{\mu\nu}=\bar{\beta}_{\mu\nu} -\frac 12 \eta_{\mu\nu}\bar{\beta}~~,
\end{equation}
\begin{alignat}{2}
	\bar{\beta}_{vv}&={r_g \over 2} \bigl(\tilde{\Phi}_\omega-\tilde{\Phi} \bigr)~~ &,~~
	\bar{\beta}_{vy}&=-r_g\omega ~\tilde{\Phi}_\omega~~, 
	\\
	\bar{\beta}_{yy}&=2r_g\omega^2 ~\tilde{\Phi}_\omega~~ &,~~
	\bar{\beta}_{za}&=0~~.
\end{alignat}
%\begin{eqnarray}\label{2.14}
%\bar{\beta}_{vv}&=&{r_g \over 2} (\tilde{\Phi}_\omega-\tilde{\Phi})~~,
%~~\bar{\beta}_{vy}=-r_g\omega ~\tilde{\Phi}_\omega~~, \\
%\bar{\beta}_{yy}&=&2r_g\omega^2 ~\tilde{\Phi}_\omega~~,
%~~\bar{\beta}_{za}=0~~.
%\end{eqnarray}
The gauge conditions are reduced to 
\begin{equation}\label{2.15a}
m^\mu\bar{\beta}_{\mu\nu}=0~~, 
\end{equation}
and allow one to find
\begin{equation}\label{2.15}
\bar{\beta}_{ua}={1 \over 2m_v}\bar{\beta}_{ab}~m^b~~,~~\bar{\beta}_{uu}={1 \over 4m^2_v}\bar{\beta}_{ab}~m^a m^b~~.
\end{equation}
Equations \eqref{2.15a} fix the components up to transformations 
${\beta}_{\mu\nu}\to {\beta}_{\mu\nu}+a_\mu m_\nu+a_\nu m_\mu$, where $a^\mu m_\mu=0$.
This  freedom corresponds to some coordinate transformations which do not change 
\eqref{i.1} and the total radiated energy \eqref{i.2}.  The arbitrariness is eliminated, if $\beta_{ab}$ are required to be unchanged. 

The subsequent analysis is based on  arguments analogous to those of \cite{Fursaev:2023lxq} for electromagnetic 
pulses caused by null strings. In coordinates $U,r,x^A$ the denominator in  \eqref{2.12} can be written as 
$$
x^\mu m_\mu +ia\varepsilon =U+r\bigl(\vec{m}\cdot \vec{n}+1\bigr)+ia\varepsilon~~,
$$
where $\vec{n}=\vec{x}/r$, $x^i={x,y,z}$, is the unit vector defined above. We are interested in an asymptotic  of \eqref{2.12} at large $r$.
The integration in \eqref{2.12} can be decomposed into two regions: a domain, where the factor $(\vec{m}\cdot \vec{n})+1$ is small,
that is, $\vec{m}$ is almost $- \vec{n}$ and the rest part of $S^2$. Let us introduce a dimensionless parameter $\Lambda$ such that
\begin{equation}\label{2.16}
{\sqrt{U^2+a^2} \over r} \ll \Lambda^2 \ll 1~~.
\end{equation}
We define the first region as $(\vec{m}\cdot \vec{n})+1\leq \Lambda^2$, and the second region as 
$(\vec{m}\cdot \vec{n})+1>\Lambda^2$
After some algebra one gets the following estimate (for components in the Minkowski coordinates) for the first region:
\begin{equation}\label{2.17}
h_{1,\mu\nu}(r,U,\Omega)\simeq  {N(\Omega) \over r}\Re \left(\tilde{\beta}_{\mu\nu}\ln
\left({U+ia\bar{\varepsilon}+r\Lambda^2 \over U+ia\bar{\varepsilon}}\right)\right)~~,
\end{equation}
\begin{equation}\label{2.18} 
\tilde{\beta}_{\mu\nu}=\beta_{\mu\nu}\mid_{\vec{m}=-\vec{n}}~~,~~
\bar{\varepsilon}=\varepsilon\mid_{\vec{m}=-\vec{n}}~~,
\end{equation}
\begin{equation}\label{2.19}
N(\Omega)=\frac{1}{4\pi^2}\sqrt{\frac{2}{1-n_x}}~~.
\end{equation}
The leading large $r$ approximation  in the second region can be easily computed
\begin{equation}\label{2.20}
h_{2,\mu\nu}(r,U,\Omega)\simeq  {b_{2,\mu\nu}(\Omega,\Lambda) \over r}~~,~~
b_{2,\mu\nu}(\Omega,\Lambda)=-C 
\int_{S^2_\Lambda} d\Omega'~\Re\left({\beta_{\mu\nu} (\Omega')\over \vec{n} \cdot \vec{m}(\Omega')+1}\right)~~.
\end{equation}
Here $S^2_\Lambda$ is a part of $S^2$ 
with the restriction $(\vec{m}\cdot \vec{n})+1>\Lambda^2$. As a result
of \eqref{2.17}, \eqref{2.20} the solution at large $r$ is a sum of a static, $h_{s,\mu\nu}$,
and a time-dependent or dynamical, $h_{d,\mu\nu}$, parts:
\begin{equation}\label{2.21}
h_{H,\mu\nu}(r,U,\Omega)\simeq h_{1,\mu\nu}+h_{2,\mu\nu}=h_{s,\mu\nu}+h_{d,\mu\nu}~~,
\end{equation}
\begin{alignat}{2}
	h_{s,\mu\nu}(r,U,\Omega) &\simeq  {b_{\mu\nu}(\Omega,\Lambda) \over r}&,~~
	b_{\mu\nu}(\Omega,\Lambda) &=N(\Omega)~\Re~ \tilde{\beta}_{\mu\nu} (\Omega)\ln (r/\varrho)+ b_{2,\mu\nu}(\Omega,\Lambda)~~, \label{2.22} 
	\\
	h_{d,\mu\nu}(r,U,\Omega) &\simeq  {H_{\mu\nu}(U,\Omega) \over r}~&,~
	H_{\mu\nu}(U,\Omega) &=N(\Omega)~ \Re\left(\tilde{\beta}_{\mu\nu}\ln
	\left({U+ia\bar{\varepsilon} \over a}\right) \right)~~, \label{2.23}
\end{alignat}
where $\varrho=a/\Lambda^2$. It is the dynamical part which contributes to the energy flux.

Although contribution from the massive source and the created dynamical perturbation enter (\ref{2.21}) on an equal footing one should keep in mind that
field of the source is proportional to the mass $m$ while the dynamical perturbation is proportional to $\omega m$.
Since $\omega$ is extremely small the perturbations caused by the string are many orders of magnitude smaller than gravitational field of the mass.

\section{Calculating the flux of gravity waves}\label{flux}
\setcounter{equation}0

\subsection{Polyhomogeneous space-times}\label{poly}

According to Eqs.~\eqref{1.13}, \eqref{1.14} the metric of space-time sourced by the mass and the straight null string  
in the linear approximation has the form \eqref{3.1}.
It has been shown in the previous section that perturbations at the future null infinity have the following asymptotics (in Minkowsky
coordinates):
\begin{equation}\label{3.2}
h^{(1)}_{\mu\nu}(r,U,\Omega)\simeq   r^{-1} \Bigl( H^{(1)}_{\mu\nu} (U,\Omega) +  \ln (r/\varrho)~ \tilde{H}^{(1)}_{\mu\nu} (\Omega) \Bigr)+
O(r^{-2}\ln r)~~~,
\end{equation} 
\begin{equation}\label{3.2a}
H^{(1)}_{\mu\nu}(U,\Omega)=H_{\mu\nu}(U,\Omega)+\hat{H}_{\mu\nu}(\Omega)~~~.
\end{equation} 
Here $H_{\mu\nu}$ is  defined in \eqref{2.23}, and $\hat{H}_{\mu\nu}$ is a static contribution
of the gravitational field of the source determined by  \eqref{1.17}. 
The gauge conditions in \eqref{1.20a} impose constraints on the amplitudes.  According to 
Eqs.~\eqref{2.22},  \eqref{2.23},  $H_{\mu\nu} (U,\Omega)$ and $\tilde{H}^{(1)}_{\mu\nu}(\Omega)$
depend on the complex tensor $\tilde{\beta}_{\mu\nu}(\Omega)$ defined in \eqref{2.18}.
Equations~\eqref{2.15a} and \eqref{2.18} imply the constraints related with the gauge conditions:
\begin{equation}\label{3.3}
\bar{m}^\mu \left( \tilde{\beta}_{\mu\nu} -\frac 12 \eta_{\mu\nu}\tilde{\beta} \right)=0~~,
\end{equation} 
where $\bar{m}$ is a past-directed null vector with components $\bar{m}^0=-1, \vec{m}=-\vec{n}$.
This vector is nothing but a null normal of the light cone $U=0$, see Fig.~\ref{Cone}. That is, $m_\mu\sim \delta^U_\mu$.
In coordinates \eqref{2.1} consequences of \eqref{2.22},  \eqref{2.23} and \eqref{3.3} are
\begin{equation}\label{3.4}
H_{r\nu} -\frac 12 \eta_{r\nu}H=0~~,~~\tilde{H}^{(1)}_{r\nu} -\frac 12 \eta_{r\nu}\tilde{H}^{(1)}=0~~,
\end{equation} 
\begin{equation}\label{3.5}
H_{rr} =H_{rA}=\gamma^{AB}H_{AB}=0~~,~~
\tilde{H}^{(1)}_{rr} = \tilde{H}^{(1)}_{rA}=\gamma^{AB}\tilde{H}^{(1)}_{AB}=0~~,
\end{equation} 
where $\gamma_{AB}$ is the metric on a unit 2-sphere.

Asymptotic \eqref{3.2}, however, is not "standard" due to the presence of the logarithmic terms with amplitude  
$\tilde{H}^{(1)}_{\mu\nu} (\Omega)$. Such space-times are called asymptotically polyhomogeneous  space-times \cite{Chrusciel:1993hx}. Their key property is in non-analytical behavior at conformal null infinities, see e.g. \cite{Godazgar:2020peu,Kroon:1998tu,ValienteKroon:1998vn,ValienteKroon:2001pc,Geiller:2022vto,Fuentealba:2022xsz,Kehrberger:2024clh,Kehrberger:2024aak}.
The logarithmic terms result in the corresponding modification of the peeling properties of  
the curvature invariants \cite{Friedrich:2017cjg}.  

To see this we can follow a standard method and calculate Newman-Penrose curvature scalars
\begin{equation}\label{rs.9}
\Psi_0 = C_{\mu\nu\rho\sigma} {\bf l}^\mu {\bf m}^\nu {\bf l}^\rho {\bf m}^\sigma~~,~~
\Psi_1 = C_{\mu\nu\rho\sigma} {\bf l}^\mu {\bf n}^\nu {\bf l}^\rho {\bf m}^\sigma~~, 
\end{equation}
\begin{equation}\label{rs.101}
\Psi_2 = C_{\mu\nu\rho\sigma} {\bf l}^\mu {\bf m}^\nu \bar{{\bf m}}^\rho {\bf n}^\sigma~~,~~
\Psi_3 = C_{\mu\nu\rho\sigma} {\bf l}^\mu {\bf n}^\nu \bar{{\bf m}}^\rho {\bf n}^\sigma ~~,~~
\Psi_4 = C_{\mu\nu\rho\sigma} {\bf n}^\mu \bar{{\bf m}}^\nu {\bf n}^\rho \bar{{\bf m}}^\sigma~~
\end{equation}
constructed out of the Weyl tensor $C_{\mu\nu\lambda\rho}$. A null tetrade is defined in coordinates \eqref{2.1} as
\begin{equation}\label{rs.10}
{\bf l}=  \frac{1}{\sqrt{2}} \partial_r~~,~~{\bf n}=  \frac{1}{\sqrt{2}} \bigl(2 \partial_U - \partial_r \bigr)~~,~~
{\bf m}= \frac{1}{\sqrt{2}r}\Bigl(\partial_\theta + \frac{i}{\sin\theta} \partial_\phi \Bigr) ~~,~~
\bar{{\bf m}}={\bf m}^*~~
\end{equation}
such that ${\bf l}^2={\bf n}^2={\bf m}^2=0$, $({\bf l}\cdot {\bf n})=-1$, $({\bf m}\cdot \bar{{\bf m}})=1$. 

"Standard" asymptotic properties imply that $\Psi_n\sim r^{n-5}$ at large $r$. However in the considered case one finds 
in the linear approximation that
\begin{equation}\label{rs.11}
\Psi_0\sim {\psi_0 \over r^3}~~,~~
\Psi_1\sim {\psi_1 \over r^3}\ln r~~,~~
\Psi_2\sim {\psi_2 \over r^3}\ln r~~,~~
\Psi_3\sim {\psi_3 \over r^3}\ln r~~,~~
\Psi_4\sim {\psi_4 \over r}~~,
\end{equation}
\begin{equation}\label{rs.12}
	\begin{split}
		\psi_0 &=\xi^A\xi^B \tilde{H}_{AB}^{(1)}~~,~~\psi_1 = \xi^A \partial_A \tilde{H}_{Ur}^{(1)}~~,~~\\
		\psi_2 &= \eth^2 \tilde{H}_{Ur}^{(1)} 
		+2 \Bigl(\tilde{H}_{Ur}^{(1)} - \tilde{H}_{UU}^{(1)} \Bigr)
		+2\Bigl(\bar{\xi}^A \xi^B - \xi^A \bar{\xi}^B\Bigr) \partial_A \tilde{H}_{UB}^{(1)}~~,~~\\
		\psi_3 & = \bar{\xi}^A \left( 2 \partial_A \tilde{H}_{UU}^{(1)}-\partial_A \tilde{H}_{Ur}^{(1)} \right)~~,~~
\psi_4  = \bar{\xi}^A \bar{\xi}^B \partial^2_U    C_{AB}  ~~,~~	
	\end{split}
\end{equation}
where $\xi=r ~{\bf m}$ and $\eth_A$ is the covariant derivative on $S^2$, and 
$C_{AB}(U,\Omega)\equiv H_{AB}(U,\Omega)$. 
The same formula for $\psi_0$ can be
found in \cite{Godazgar:2020peu}. Asymptotics of $\Psi_0,\Psi_1,\Psi_2$ also correspond to results of \cite{Gasperin:2017apb}.

The logarithmic terms in \eqref{3.2} result in static tidal forces. For instance, a test massive particle, which is at rest
at a distance $r$ with respect 
to the considered source, acquires an additional constant coordinate acceleration with components
$\delta w^r\sim r^{-2} \ln r~ \tilde{H}_{UU}$, $\delta w^A\sim r^{-3} \ln r ~(\tilde{H}_{UU})^{,A}$.  Hence, 
a distance $L$ between two nearby test particles changes after a time $T$ by $\delta L \sim L~ r^{-3}\omega r_g T^2 \ln r$.

\subsection{Metric perturbations as gravity waves}\label{strain}

Our interest is in the dynamical (time-dependent) part $h_{d,\mu\nu}(r,U,\Omega)\simeq   r^{-1} H_{\mu\nu} (U,\Omega)$ of metric perturbations in \eqref{3.2}. Gauge conditions in \eqref{1.20a} are invariant under residual coordinate transformations 
\begin{equation}\label{rs.1}
\delta x^\mu=\zeta^\mu(x)~~,~~\Box \zeta^\mu=0~~.
\end{equation}
These transformations can be used to impose futher restrictions on $H^{(1)}_{\mu\nu}$. We assume the following asymptotic 
form, in the Minkowsky coordinates:
\begin{equation}\label{rs.2}
\zeta_\mu(x)\sim r^{-1}(\sigma_\mu(U,\Omega)+\tilde{\sigma}_\mu(U,\Omega)\ln r)+O(r^{-2}\ln r)~~.
\end{equation}
One can show that equations $\Box \zeta_\mu=0$ imply only that $\partial_U\tilde{\sigma}_\mu=0$, and
subleading terms in \eqref{rs.2} are determined by $\sigma_\mu$ and $\tilde{\sigma}_\mu$.  Calculations also show that 
\eqref{rs.1} generate transformations of the metric perturbations $(h')^{(1)}_{\mu\nu}=h^{(1)}_{\mu\nu}+\nabla_\mu \zeta_\nu
+\nabla_\nu \zeta_\mu$ so that asymptotic form of $(h')^{(1)}_{\mu\nu}$ coincides with \eqref{3.2} where
the components of the dynamical part now look as 
\begin{equation}\label{rs.3}
H'_{r\mu}=H_{r\mu}~~,~~H'_{AB}=H_{AB}~~,
\end{equation}
\begin{equation}\label{rs.4}
H'_{UU}=H^{(1)}_{UU}+2\partial_U\sigma_U~~,~~H'_{Ur}=H_{Ur}+\partial_U\sigma_r~~,~~
H'_{UA}=H^{(1)}_{UA}+\partial_U\sigma_A~~.
\end{equation}
Therefore the residual transformations can be used to impose additional conditions on the dynamical part 
\begin{equation}\label{rs.5}
H_{U\mu}=0~~.
\end{equation}
By taking into account \eqref{3.5} one can readily conclude that the only non-vanishing components are spherical components, 
$C_{AB}(U,\Omega)$, 
which appear in \eqref{i.1}. In the Minkowsky coordinates $t,x^i$ the non-zero components
of the dynamical part of perurbation \eqref{3.2} can be written as
\begin{equation}\label{rs.6}
h_{d,ij}(r,U,\Omega)\simeq   r^{-1} H_{ij} (U,\Omega)~~,~~H_{ij}=y^A_{,i}y^B_{,j}C_{AB}~~,~~
\end{equation}
\begin{equation}\label{rs.7}
H_{ij}\delta^{ij}=0~~,~~x^iH_{ij}=0~~,
\end{equation}
where $y^A=(\theta,\varphi)$. Time componets of the perturbation vanish as a result of \eqref{rs.5}. 
The dynamical perturbations are spatially traceless and othogonal to the direction of motion set by vector $x^i$. 

Thus, traceless tensors $H_{ij}$ or $C_{AB}$ 
can be interpreted as amplitudes of outgoing gravitational waves. Constraints \eqref{rs.7}
leave two independent degrees of freedom which can be related to two polarizations. By doing decomposition of $H^{(1)}_{ij}$
over the spherical harmonics along the lines of \cite{Thorne:1980ru} one gets a multipole expansion of the gravitational radiation, including a quadrupole term.

A coordinate invariant nature of this conclusion can be seen if we use the Bel decomposition and calculate asymptotic of the electrogravic
(or tidal) tensor 
\begin{equation}\label{rs.8}
E_{\mu\nu}=R_{\mu\lambda\nu\rho}u_o^\lambda u_o^\rho~~,
\end{equation}
where $u_o$ is a 4-velocity vector of a distant observer. We assume this observer is at rest.   By taking into account that in the considered case the Riemann and Weyl tensors
coincide one finds the following expressions for invariants constructed out of $E_{\mu\nu}$ with the help of tetrades ${\bf e}_p$, 
defined in \eqref{rs.10},
\begin{equation}\label{rs.13a}
E_{ll}=E_{nn}=-E_{ln}=E_{m\bar{m}}=\Psi_2+\Psi_2^*~~,
\end{equation}
\begin{equation}\label{rs.13b}
E_{mm}=\Psi_0+\Psi_4^*~~,~~
-E_{lm}=E_{nm}=\Psi_1-\Psi_3^*~~,
\end{equation}
where $E_{pq}=E_{\mu\nu}{\bf e}_p^\mu {\bf e}_q^\nu$.  According to 
\eqref{rs.11}, \eqref{rs.12} at future null infinity
\begin{equation}\label{rs.14}
E_{mm}\sim \Psi_4^*\sim {\xi^A\xi^B \partial^2_U    C_{AB}  \over r} ~~,
\end{equation}
while other non-vanishing components of $E_{pq}$ decay faster than $1/r$. Thus, the strain tensor $C_{AB}$ of the gravity wave
determines the leading asymptotic of the electrogravic tensor. This also determines the leading term 
\begin{equation}\label{rs.15}
E_{\mu\nu}E^{\mu\nu}\sim {2 \over r^2} \partial^2_U    C_{AB}  \partial^2_UC^{AB}+O(r^{-4})~~,
\end{equation}
in the so called super-energy scalar.

\subsection{Going beyond the linearized approximation}\label{set}

We analyze now the energy carried away by gravity waves in case of this sort of polyhomogeneous space-times. By following 
the standard approach \cite{Misner:1973prb},\cite{Landau2},\cite{Wald:1984rg}, to this aim one needs to go beyond 
the linearized approximation and write 
\begin{equation}\label{3.7}
g^{\mbox{\tiny{out}}}_{\mu\nu}\simeq \eta_{\mu\nu}+h^{(1)}_{\mu\nu}+h^{(2)}_{\mu\nu}~~,
\end{equation} 
instead of \eqref{3.1}. Second order correction $h^{(2)}_{\mu\nu}$ is determined by the Einstein equations
\begin{equation}\label{3.8}
G_{\mu\nu}=\frac 12 \kappa T_{\mu\nu}~~,
\end{equation} 
where $T_{\mu\nu}$ is the stress-energy tensor of the source which first appeared in \eqref{1.16}. By taking into account 
the linearized equations for $h^{(1)}_{\mu\nu}$ one gets from \eqref{3.8}
\begin{equation}\label{3.9}
G^{(1)}_{\mu\nu}(h^{(2)})=-G^{(2)}_{\mu\nu}(h^{(1)})\equiv \frac 12 \kappa t_{\mu\nu}~~,
\end{equation}
where $G^{(k)}$ is a term in decomposition of the Einstein tensor $G_{\mu\nu}$ over the flat metric which contains
$k$-th power of $h^{(1)}$. 
The quantity $t_{\mu\nu}$ is the pseudo-tensor of the gravitational field. It is divergence free,
$\partial^\mu t_{\mu\nu}=0$,  since $\partial ^\mu G^{(1)}_{\mu\nu}=0$. The energy flow is determined by $t^r_U$.

We proceed with computations in the harmonic gauge, 
$\partial^\mu \bar{h}^{(1)}_{\mu\nu}=\partial^\mu \bar{h}^{(2)}_{\mu\nu}=0$. Then the left and right sides of \eqref{3.9} are:
\begin{multline}\label{3.10} 
	(G^{(1)})^r_U(h^{(2)})=-\frac 12 \Box (h^{(2)}) ^r_U= 
	\\
	=-\frac 12\left(\partial_r^2-2\partial_U \partial_r -{2 \over r} (\partial_U-\partial_r)+ {1 \over r^2}(\eth_A \eth^A-2)\right)(h^{(2)}) ^r_U
	+{1 \over r^3}\eth^A h^{(2)}_{UA}~~, 
\end{multline}
\begin{equation}\label{3.11}
	-(G^{(2)})^r_U(h^{(1)})=\frac 14 \Bigl( \partial_U(h^{(1)})^{AB}\partial_U(h^{(1)})_{AB}-
	\partial^2_U \bigl(  (h^{(1)})^{AB}(h^{(1)})_{AB}  \bigr)  \Bigr)~~. 
\end{equation}
We now suppose that $h^{(2)}$ has asymptotic behavior 
\begin{multline} \label{3.12}
	h^{(2)}_{\mu\nu}(r,U,\Omega)\simeq   r^{-1} \Bigl( H^{(2)}_{\mu\nu} (U,\Omega) +  \ln (r/\varrho)~ 
	\tilde{H}^{(2)}_{\mu\nu} (U,\Omega)+ 
	\\
	+ \ln^2 (r/\varrho)~ 
	\tilde{Q}^{(2)}_{\mu\nu} (U,\Omega)  \Bigr)+
	O(r^{-2}\ln^2 r)~~.
\end{multline}
The differences between \eqref{3.2} and asymptotic \eqref{3.12} are the following:
i) we admit possibility of higher order powers of the logarithms \cite{Godazgar:2020peu,Kroon:1998tu,ValienteKroon:1998vn,ValienteKroon:2001pc,Geiller:2022vto,Fuentealba:2022xsz},
ii) we do not require that amplitudes $\tilde{H}^{(2)}$, $\tilde{Q}^{(2)}$
are static. Also note that conditions \eqref{3.4}, \eqref{3.5} are not to be applicable to $H^{(2)}$, $\tilde{H}^{(2)}$,
$\tilde{Q}^{(2)}$.  One can show that the harmonic gauge conditions imply only time independence
of some components
\begin{equation}\label{3.12a}
\partial_U H^{(2)}_{r\mu}=\partial_U \tilde{H}^{(2)}_{r\mu}=\partial_U \tilde{Q}^{(2)}_{r\mu}=0~~~,~~\mu=r, A~~,
\end{equation}
\begin{equation}\label{3.12b}
\partial_U (\gamma^{AB} H^{(2)}_{AB})=\partial_U (\gamma^{AB} \tilde{H}^{(2)}_{AB})=
\partial_U (\gamma^{AB} \tilde{Q}^{(2)}_{AB})=0~~~.
\end{equation}
Substitution of \eqref{3.2} in \eqref{3.11} and \eqref{3.12} in \eqref{3.10} yields:
\begin{equation}\label{3.13}
(G^{(1)})^r_U(h^{(2)})\simeq {1 \over r^2} \partial_U (\tilde{H}^{(2)}) ^r_U +
{2 \over r^2} \partial_U (\tilde{Q}^{(2)}) ^r_U ~\ln (r/\varrho)
+O(r^{-3}\ln^2 r)
~~,
\end{equation}
\begin{equation}\label{3.17}
t^r_U \simeq {2 \over \kappa r^2} \Bigl(N_{AB} N^{AB} 
-\partial_U(C_{AB}N^{AB})- \partial_U(\tilde{H}^{(1)}_{AB}~ N^{AB}) \ln (r/\varrho)\Bigr)+O(r^{-3}\ln^2 r)~~,
\end{equation}
where $N_{AB}\equiv \frac 12 \partial_UC_{AB}$ and indices $A,B$ are now risen with the metric on unit $S^2$. From Eq.~\eqref{3.9} one
comes to relations between 1-st order and 2-d order perturbations in \eqref{3.7}
\begin{equation}\label{3.17a}
\partial_U (\tilde{H}^{(2)}) ^r_U = -N_{AB} N^{AB} +\partial_U(C_{AB}N^{AB})~~,
\end{equation}
\begin{equation}\label{3.17b}
2\partial_U (\tilde{Q}^{(2)}) ^r_U= \partial_U(\tilde{H}^{(1)}_{AB}~ N^{AB})~~.
\end{equation}
Equation~\eqref{3.17b} justifies inclusion the $\ln^2 r$-term in \eqref{3.12}. It is interesting to note that in the harmonic gauge 
only the logarithmic terms in $h^{(2)}_{\mu\nu}$ are responsible for the flux of gravity waves.

\subsection{Intensity of the gravitational radiation}\label{inten}

The stress-energy pseudo-tensor $t^\mu_\nu$ depends on the chosen gauge and it does not describe any localized energy.
The physical meaning can be attributed to the global energy obtained by integrating $t^0_0$ over a space-like hypersurface.
This integrated quantity should be invariant under coordinate transformations  which preserve the leading asymptotic of the metric,
see e.g. \cite{Wald:1984rg}.  We demonstrate this property in Appendix~\ref{app2} for $t^r_U$ component.

Consider the integrated energy in a ball $r\leq R$ of sufficiently large radius $R$. The change of the energy with time 
is given by the energy flux through the spherical boundary
\begin{equation}\label{3.18}
W(U,R)= \int_{r=R} d\Omega~ r^2~ t^r_U(r,U,\Omega)~~.
\end{equation}
This quantity is certainly not finite in the limit $R\to\infty$ due to the presence of the logarithmic term in r.h.s. of \eqref{3.17}.
To understand whether this divergence is physically relevant one needs to calculate the energy 
\begin{equation}\label{3.19}
E(U_2,U_1,R)=\int_{U_1}^{U_2} dU~ W(U,R)~~,
\end{equation}
emitted over a period of time $U_2-U_1\equiv \Delta U$ ($U_2>0$, $U_1<0$) much larger than a typical wave-length of the system. Since the impact parameter $a$  can be interpreted 
as a "size" of the system (the string and the mass), which determines a duration of the gravitational pulse and an effective wave-length of the emitted radiation, one needs $\Delta U\gg a$.

When integrating  $t^r_U $ over $U$ contributions of the last two terms in r.h.s. of \eqref{3.17} 
are suppressed at $|U|\gg a$ as a result of the following asymptotics:
\begin{equation}\label{3.19a}
C_{AB}N^{AB}\sim U^{-1}\ln (|U|/a)~~,~~\tilde{H}^{(1)}_{AB}~ N^{AB}~\ln (r/\varrho) \sim U^{-1}\ln (r/\varrho)~~.
\end{equation}
Let us remind that our approximation is obtained under assumption \eqref{2.16} which implies that $|U| \ll R$. If one puts
$U_i=c_iR$, where $c_i$ are some constants, $|c_i|\ll 1$, $c_2>0$, $c_1<0$, and takes into account \eqref{3.19a} the total radiated energy $E$ defined below does not depend on the logarithmic
terms at large $R$ and takes the form
\begin{equation}\label{3.20}
E=\lim_{R\to\infty} E(c_2R,c_1R,R)= {2 \over \kappa} \int_{-\infty}^{\infty} dU \int d\Omega~ N_{AB} N^{AB}~~.
\end{equation}
We conclude that the logarithmic divergence in the flux disappears when averaging the radiated energy on time intervals much larger 
than the impact parameter $a$. So the divergence is irrelevant.

In a similar way we define an averaged intensity of the flux 
as
\begin{equation}\label{3.22}
I(\Omega)\equiv {1 \over a}\lim_{R\to \infty} \int_{c_1R}^{c_2R} dU~ R^2~ t^r_U(R,U,\Omega) =
{2 \over a\kappa} \int_{-\infty}^\infty dU~ N_{AB} N^{AB}~~.
\end{equation}
Relation between the averaged intensity and the total emitted energy is
\begin{equation}\label{3.23}
E=a \int d\Omega~ I(\Omega)~~.
\end{equation}
Straightforward but tedious computations presented in Appendix~\ref{App1} yield:
\begin{equation}\label{3.24}
I(\Omega)  
	=
	\frac{\pi N(\Omega)^2}{4 \kappa \bar{\varepsilon}(\Omega)}
	\left( \alpha (\Omega)+ \frac{\gamma(\Omega)}{a^2 \bar{\varepsilon}^2(\Omega)} \right)~~,
\end{equation}
where $N$ is defined in \eqref{2.19}, $\bar{\varepsilon}$ in \eqref{a1.1}  and $\alpha$, $\gamma$ in \eqref{a1.2}, \eqref{a1.4}.

\begin{figure}
	\begin{center}
	\includegraphics[width=8cm]{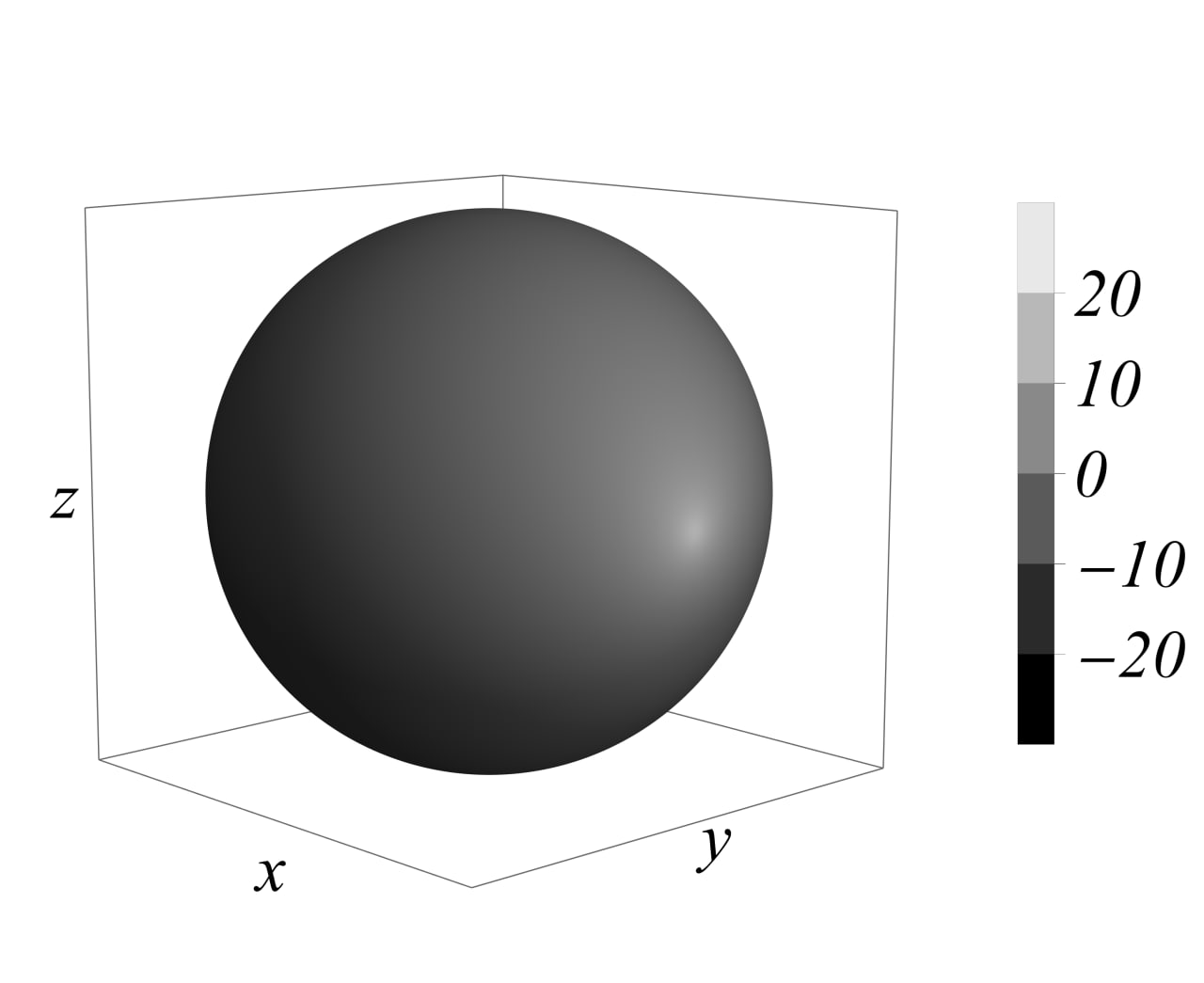}
	\end{center}
	\caption{shows the angular distribution of the logarithm of the averaged intensity of the generated gravitational radiation. The intensity is concentrated in the direction of the string motion, on the axis $y=z=0$.}
\label{f2}
\end{figure}

Angular distribution of the logarithm of the intensity is shown on Fig. \ref{f2}.  The picture looks similar to the case of electromagnetic radiation generated by a straight null string and an electric of magnetic-dipole point-like source. The intensity is concentrated near the line which goes through the source parallel to the velocity of the string and in the direction of the string motion.

Let us emphasize that we present $I(\Omega)$ just to give a qualitative picture of the angle distribution of the gravitational radiation.  
We do not discuss in the present work more physically interesting characteristics, such as, for example, response of gravitational detectors to string generated gravitational  waves. 

\subsection{Deformation of the string and displacement of the source}\label{rob}

The results obtained for metric perturbations imply that position of the source is fixed and the string is straight. In reality, the source and each point of the string move along their geodesics in a weak gravitational field.  Changes in the position of the source and in the form of the string may affect the analysis given above. To estimate possible corrections, we assume 
that each small section of the string contributes independently to the metric perturbation. 

Let $a_0+\delta a$ be an impact parameter  
of a given section, where $a_0$ is an impact parameter of the section for straight string and initial position of the source.
An addition $\delta a$ appears due to the deformation of the string and the displacement of the source. If $\delta a/a_0$ is small for each segment of the string, then correction to the perturbation of the metric should be
suppressed by factor $O(\max(\delta a/a_0))$ as compared to the main effect.

The interaction of the string with the source takes an infinite time, so $\delta a$ even in the weak gravitational field of the source can be infinitely large. However, the flux of gravitational radiation, determined by $N_{AB}N^{AB}$, describes the gravitational burst 
at moments $U\sim O(a)$. Since 
$\delta a$ during this time are expected to be of order $r_g$ (see explicit calculations in \cite{Davydov:2022qil}), the relative correction for the contribution of the each section of the string, and, accordingly, of the entire string, is suppressed by factor $O(r_g/a)$. This effect is small enough to be neglected.

As for the displacement of the source under the action of the gravity wave, the acceleration of the source can be estimated as $\omega r_g/a^2$. During a time interval $\Delta U=O(a)$ the source may be moved at a distance not larger than $O(\omega r_g)$.
This displacement is extremely small and does not affect our results as well.

\subsection{Averaged luminosity and spin effects}\label{exp}

Let us  estimate now the averaged luminosity of the flux, which can be crudely defined as:
\begin{equation}\label{4.1}
\dot{E}\equiv {E \over a}\sim ~C ~\frac{\omega^2}{a^2}~ \frac{r_g^2}{G}~~,
\end{equation}
where $C$ is a numerical coefficient, $C< O(10^{-2})$ and $r_g=2MG$ is the gravitational radius of the source.
For tensile cosmic strings with tension $\mu$ the CMB spectrum yields the constraint $G\mu \leq 10^{-7}$, see
\cite{Planck:2013mgr,Charnock:2016nzm,Dvorkin:2011aj}, while the stochastic gravitational-wave background 
gives a stronger limit, $G\mu \leq 10^{-12}$, see \cite{LIGOScientific:2021nrg}. 

Since we are dealing with gravitational radiation the  parameter $\omega_0=1\cdot  10^{-12}$ can be used as some reference value,
and \eqref{4.1} can be rewritten as 
\begin{equation}\label{4.2}
\dot{E} \sim ~C \left({\omega \over \omega_0}\right)^2 \left({r_g \over a}\right)^2 \cdot 10^{36}~{\mbox{erg} \over \mbox{s}}~~.
\end{equation}
Thus, a null string and a massive source with impact parameter $a=10~r_g$ and $\omega \sim \omega_0$
have the luminosity of the order $10^{34}$ erg per second, which is comparable with gravity wave luminosity of such systems as the double 
pulsar PSR B1913+16. Yet such luminosity is extremely smaller than the peak luminosity of gravitational radiation produced by 
binary black hole mergers  (about $10^{56}$ erg per second).

If the source has a spin $J_j$ there will be non-trivial components in Eq.~\eqref{1.17}
\begin{equation}\label{1.17s}
\hat{h}_{0i}=4G~\varepsilon_{ijk}J_j\partial_k\phi~~,
\end{equation}
where $i,j,k$ correspond $x,y,z$.  These components yield additional contributions to string generated gravity waves.
As a result one expects two additions to \eqref{4.2}: one is due to the radiated energy $E_{mJ}$ proportional to 
mass $m$ of the source and spin $J$,  another, $E_{J}$, is of the second order in $J$.
By taking into account that spin factors may appear in the dimensionless combination $GJ/a^2$ it is easy to come to the 
following estimates:
\begin{align}
	\dot{E}_{mJ}  &\sim ~\dot{E}   \left({a \over r_g}\right) \left({GJ \over a^2}\right)~~, \label{4.3}
	\\
	\dot{E}_{J} &\sim ~  \dot{E} \left({a \over r_g}\right)^2 \left({GJ \over a^2}\right)^2~~.
	\label{4.4}
\end{align}
Suppose the source is a Kerr black hole. In this case one can introduce
the black hole parameter $a_{bh}=J/M$, such that the black hole is extremal if $a_{bh}=MG$.
Relations  \eqref{4.3}, \eqref{4.4} can be transformed as follows:
\begin{align}
	\dot{E}_{mJ}  &\sim ~\dot{E}   \left({a_{bh} \over a}\right) ~~, \label{4.5}
	\\
	\dot{E}_{J} &\sim ~   \dot{E}  \left({a_{bh} \over a}\right)^2~~. \label{4.6}
\end{align}
Thus, contributions to the  luminosity from the spin are suppressed by factors of $a_{bh}/ a < r_g/a$, as compared to the luminosity
due to the mass of the source. One can neglect spin effects in crude estimates of the luminosity.

\subsection{Detection of string generated gravity waves}\label{SGWB} 

As it follows from Eqs.~\eqref{2.21}-\eqref{2.23} the amplitude of a gravitational
wave outgoing from the string and a source has the order of magnitude 
$h\sim \omega r_g/r$. Earth-based gravitational antenna have the sensitivity
to gravity waves with wave lengths in the interval $50-500~$km and amplitude $\sim 10^{-21}$.
Thus, the impact parameter $a$ of our system has to be of the same order, and the most suitable sources
to produce gravity waves from moving null strings are stellar mass black holes with  $r_g\sim 1 ~ $km. For $\omega =O(10^{-12})$ and a black hole located in a nearby galaxy, say, in the Andromeda Galaxy (M31), with the distance to
the Earth  about $r\sim 1~\mathrm{Mpc}\sim 10^{19}~$km,
the amplitude of the gravity wave would be $\sim 10^{-31}$, too small to be detected. Gravity waves can be also observed by 
possible space-based interferometers. For supermassive black holes  the waves generated 
by null strings have much larger amplitudes, but they are still very small to be observed.
Therefore,  direct registration of signals produced by null cosmic strings from massive sources is a challenging task. 

If exist, the null cosmic strings, when moving through the matter distributed 
over the Universe, produce numerous gravity waves of different wave lengths at different epochs.
All these waves contribute to a stochastic gravitational wave background (SGWB).
Detection of SGWB and  understanding its sources is one of the important purposes of gravitational wave experiments, see   
\cite{Caprini:2018mtu}  for a review.

In addition to compact massive sources, such as black holes and neutron stars, the Universe may contain clumps of cold dark matter (DM). Within existing models~\cite{Kashiyama:2018gsh,Brax:2020oye}, small clumps can have masses up to thousands of solar masses and sizes up to few parsecs, with a diverse density profiles, from a density of the order of $M_{\odot}\,\mathrm{pc}^{-3}$ to a density in the core comparable to the neutron star density. Detection of such small clumps by microlensing is difficult, so the possibilities of detecting gravitational wave signals from clumps are being actively explored, both using detectors such as LISA, LIGO~\cite{Jaeckel:2020mqa}, and by using pulsar timings~\cite{Kashiyama:2018gsh}.

Gravitational responses of detectors or pulsars when a clump of DM passes nearby, as well as GW effects caused by clump fragmentation can be significant~\cite{Chatrchyan:2020pzh}.  The above analysis of string generated gravity waves from point sources can be generalized to the case of GW from compact or fairly sparse clumps of DM. Since such waves should contribute to SGWB, their effects may provide another theoretical possibility for detecting the dark matter.

Below we  estimate gravitational effects from a string and a DM clump of mass $M$ which is uniformly distributed over a spatial domain with a characteristic size  $L$. As earlier, we assume that the string is straight.
By using \eqref{2.23} and assuming that $r\gg L$  one gets for the news tensor
\begin{equation}\label{4.7}
N_{AB}\simeq N(\Omega)~\Re \left(\tilde{\beta}_{AB}(\Omega)\int_{a}^{a+L}  {\rho~dy \over U(y)+i y\bar{\varepsilon}}\right)~~.
\end{equation}
Here integration goes over a section of the clump by the plane of constant $x$ and $z$, $\rho\sim M/L$ is
a mass density of this section, and  $a$ now is a least impact parameter between points of the clump and the string.  It follows from \eqref{4.7} that the parameter $r_g/a$, which determines the amplitude of the news tensor for a point-like mass, should be replaced 
with the ratio $GM/L$, if the clump is sufficiently homogeneous and it does not cross the string. This means that 
for the case of the clump
our estimate \eqref{4.1} should be replaced with
\begin{equation}\label{4.8}
	\dot{E}\equiv {E \over L}\sim ~C \left({\omega \over \omega_0}\right)^2 \left({r_g \over L}\right)^2 \cdot 10^{36}~{\mbox{erg} \over \mbox{s}}~~.
\end{equation}
For $M\sim 10^3M_\odot$, $L\sim 1$~pc, $\omega\sim \omega_0$, one has 	$\dot{E}\sim 10^{15}$
erg per second. Since the effective frequency of the radiation with such parameters 
is in the nanohertz range, its detection is possible by using pulsar timings. 

\section{Concluding remarks}\label{sum}
\setcounter{equation}0

The aim of this work was two-fold. First, we were going to check if the description of field effects in null-string space-times \cite{Fursaev:2022ayo}
can be  extended to weak gravitational fields. Second, we were looking for gravitational effects which could provide potential observational 
signatures of null cosmic strings.

To our knowledge, our result is the first solution in the linearized approximation for the space-time geometry sourced by a point mass and a null string. The solution is not 
stationary and asymptotically non-trivial. Since it belongs to a class of polyhomogeneous space-times calculating the flux of outgoing gravity waves requires some care.
As has been shown non-analytic terms in the  flux  can be eliminated by  averaging over a sufficiently large period of time.

One difficulty, which appears when dealing with polyhomogeneous space-times, is
that the Bondi analysis of the gravitational radiation based on the mass aspect, see e.g. \cite{Madler:2016xju}, cannot be applied here straightforwardly.  It should be emphasized that coordinates $x^\mu=(U,r,\Omega)$ we use here and the Bondi-Sachs coordinates are different and should not be confused. We work in the harmonic gauge,
while  the Bondi-Sachs coordinates are determined by the conditions $g_{rr}=g_{rA}=0$ and $\det g_{AB}=r^4\sin^2\theta$. 
These conditions hold only for $h^{(1)}$ up to $O(r^{-2})$ terms , see \eqref{3.5}, and they are violated for higher order perturbations starting with $h^{(2)}$.  It would be interesting 
to identify the mass aspect function for the considered geometry.

The luminosity generated by null strings in the form of  gravitational radiation is much higher  than the luminosity generated in the form of electro-magnetic (EM) radiation. For example, according to  \eqref{4.2}, for the string energy parameter $\omega\sim 10^5 \omega_0$
the GW luminosity of a null string and a pulsar is $10^{46}$ erg per second
against $10^{28}$ erg per second of the EM luminosity of the same system, see
\cite{Fursaev:2023lxq}.  Thus, we expect that the generated gravitational radiation can be important source of data to identify possible signatures of null cosmic strings. We are going to proceed with a more careful analysis of contribution from null strings to SGWB, by 
analogy with contribution from tensile cosmic strings.

As we have mentioned in Section~\ref{exp} spin effects related to the rotation of the source can be neglected in the averaged
luminosity.  In case of EM radiation generated by a null string near a magnetic-dipole source the intensity depends 
on orientation of the magnetic moment of the source \cite{Fursaev:2023lxq}.
Thus, one may expect that intensity of gravity waves may be sensitive to mutual orientation of the spin of the source 
and velocity of the string. It would be interesting to study these effects in a separate work.

Our arguments in Section~\ref{rob} to demonstrate that the obtained results are robust have been based  on assumption that each segment of the string contributes independently to the flux. There is one effect, which cannot be described in terms of individual contributions of string sections. This happens when a null string forms a caustic behind the source at a distance $a^2/2r_g$, see \cite{Davydov:2022qil}. Since a large amount of energy 
is concentrated in a domain around the caustic, it may create its own gravity waves. If this effect does exist its properties are expected  to be quite different from gravity waves we consider in this work.

\section{Acknowledgments}

This research is supported by Russian Science Foundation grant No. 22-22-00684.

\newpage
\appendix

\section{Computations}\label{App1}
\setcounter{equation}0

Here we give some details of how explicit form of asymptotic \eqref{3.2} looks like.
One starts with Eq.~\eqref{2.12} where tensor $\beta_{\mu\nu}$ has the following components:
\begin{equation} \label{phys_metric}
	\begin{split}
		\beta_{uu} &= \frac{r_g}{2} \biggl( \left( \frac{m_u+\omega m_y }{m_v} \right)^2 \tilde{\Phi}_\omega - \left( \frac{m_u}{m_v} \right)^2 \tilde{\Phi} \biggr)~~, \\
		\beta_{uv} &= \frac{r_g}{2}~ \omega^2  \tilde{\Phi}_\omega~~, \\
		\beta_{uy} &= r_g ~\omega ~\frac{m_u+\omega m_y }{m_v}  \tilde{\Phi}_\omega~~, \\
		\beta_{vv} &= \frac{r_g}{2} \bigl(\tilde{\Phi}_\omega - \tilde{\Phi} \bigr)~~,\\
		\beta_{vy} &= -r_g \omega \tilde{\Phi}_\omega~~,\\
		\beta_{yy} &=  r_g \biggl( - \frac{(m_u+\omega m_y - \omega^2 m_v) }{m_v} \tilde{\Phi}_\omega +  \frac{m_u}{m_v} \tilde{\Phi} \biggr)~~,\\
		\beta_{zz} &=r_g \biggl( - \frac{(m_u+\omega m_y + \omega^2 m_v) }{m_v}  \tilde{\Phi}_\omega +  \frac{m_u}{m_v}  \tilde{\Phi} \biggr)~~,\\
		\beta_{uz} &= \beta_{vz} = \beta_{yz} =  0~~.
	\end{split}
\end{equation}
Amplitude $H^{(1)}_{\mu\nu}$ is determined from  Eq.~\eqref{2.23} as 
\begin{equation}\label{a.1}
H^{(1)}_{\mu\nu}(U,\Omega) =N(\Omega)~ \Re\left(\tilde{\beta}_{\mu\nu}\ln
\left({U+ia\bar{\varepsilon} \over a}\right) \right)~~. 
\end{equation}
As has been explained in Sec.~\ref{elrad} tensor structures $\tilde{\beta}_{\mu\nu}$ are values of $\beta_{\mu\nu}$
when null vector $m^\mu$ is a normal vector to the null cone $U=0$, see \eqref{2.18}. A straightforward computation yields
\begin{equation} \label{perturb}
	\begin{split}
	\tilde{\beta}_{uu} &= \frac{r_g}{2} \biggl( \left( \frac{1+n_x-2\omega n_y }{1-n_x} \right)^2 \tilde{\phi}_\omega - \left( \frac{1+n_x}{1-n_x} \right)^2 \tilde{\phi} \biggr)~~, \\
	\tilde{\beta}_{uv} &= \frac{r_g}{2} \omega^2  \tilde{\phi}_\omega~~, \\
	\tilde{\beta}_{uy} &= r_g \omega \frac{1+n_x-2\omega n_y }{1-n_x}  \tilde{\phi}_\omega~~, \\
	\tilde{\beta}_{vv} &= \frac{r_g}{2} \bigl(\tilde{\phi}_\omega - \tilde{\phi} \bigr)~~,\\
	\tilde{\beta}_{vy} &= -r_g \omega \tilde{\phi}_\omega~~,\\
	\tilde{\beta}_{yy} &=  r_g \left( - \frac{(1+n_x-2\omega n_y - \omega^2 (1-n_x)) }{1-n_x} \tilde{\phi}_\omega +  \frac{1+n_x}{1-n_x} \tilde{\phi} \right)~~,\\
	\tilde{\beta}_{zz} &=r_g \left( - \frac{(1+n_x-2\omega n_y + \omega^2 (1-n_x)) }{1-n_x}  \tilde{\phi}_\omega +  \frac{1+n_x}{1-n_x}  \tilde{\phi} \right)~~,\\
	\tilde{\beta}_{uz} &= \tilde{\beta}_{vz} = \tilde{\beta}_{yz} =  0~~,
	\end{split}
\end{equation}
where $\tilde{\phi} = \tilde{\phi}_{\omega=0}$ and
\begin{equation}
	\tilde{\phi}_\omega = \frac{\sqrt{2} (1-n_x)^{3/2} }{ \bar{\varepsilon} \bigl( -n_y+\omega(1-n_x)+i \bar{\varepsilon}\bigr) }~~,
\end{equation}
\begin{equation} \label{a1.1}
	\bar{\varepsilon} =  \sqrt{2(1-n_x)-n_y^2}~~.
\end{equation}
If the energy of the string is small, $\omega \ll 1$,
\begin{equation}
	\begin{split}
		\tilde{\beta}_{uu} &\simeq -r_g \omega \frac{1+n_x}{2\sqrt{2}(1-n_x)^{3/2}} \left( \frac{1}{\bar{\varepsilon}} \Bigl(n_x^2-1+n_y^2(n_x-3)\Bigr)+i n_y(n_x-3)\right)+ O(\omega^2)~~, \\
		\tilde{\beta}_{uv} &\simeq 0 + O(\omega^2)~~, \\
		\tilde{\beta}_{uy} &\simeq -r_g \omega \frac{1+n_x}{\sqrt{2(1-n_x)}}\left( \frac{n_y}{\bar{\varepsilon}} + i\right) + O(\omega^2)~~, \\
		\tilde{\beta}_{vv} &\simeq r_g \omega \sqrt{\frac{1-n_x}{2}} \left( \frac{\bar{\varepsilon}^2-n_y^2}{4 \bar{\varepsilon}} - i \frac{n_y}{2} \right) + O(\omega^2)  ~~,\\
		\tilde{\beta}_{vy} &\simeq r_g \omega \sqrt{\frac{1-n_x}{2}} \left( \frac{n_y}{\bar{\varepsilon}} + i\right) + O(\omega^2)~~,\\
		\tilde{\beta}_{yy} &\simeq -r_g \omega \sqrt{\frac{1-n_x}{2}} \left( \frac{1}{\bar{\varepsilon}} \bigl(n_y^2+n_x+1  \bigr) + i n_y \right)+ O(\omega^2) ~~,\\
		\tilde{\beta}_{zz} &\simeq -r_g \omega \sqrt{\frac{1-n_x}{2}} \left( \frac{1}{\bar{\varepsilon}} \bigl(n_y^2+n_x+1  \bigr) + i n_y \right)+ O(\omega^2)~~,\\
		\tilde{\beta}_{uz} &= \tilde{\beta}_{vz} = \tilde{\beta}_{yz} =  0~~.
	\end{split}
\end{equation}
The averaged intensity of the radiation is
\begin{equation}
I(\Omega) = \frac{2}{a \kappa} \int_{-\infty}^{\infty} dU f_G(U,\Omega)~~,
\end{equation}
\begin{equation}
f_G(U,\Omega)\equiv N_{AB} N^{AB}~~,~~N_{AB} = \frac{1}{2}  \partial_U C_{AB}~~.
\end{equation}
After some algebra, by using gauge conditions \eqref{3.3}, one finds a useful expression in terms of components 
of  $N_{\mu\nu} = \frac{1}{2}  \partial_U H^{(1)}_{\mu\nu}$ in the Minkowski coordinates:
\begin{multline}
	f_G(U,\Omega) = 
	A_1 N_{vv}^2 
	+ A_2 N_{vy}^2 
	+ A_3 N_{yy}^2 
	+ A_4 N_{zz}^2 
	+ A_5 N_{vv} N_{vy} +
	\\ 
	+ A_6 N_{vv} N_{yy}
	+ A_7 N_{vv} N_{zz}  
	+ A_8 N_{vy} N_{yy} 
	+ A_9 N_{vy} N_{zz}
	+ A_{10} N_{yy} N_{zz}~~, 
\end{multline}
\begin{alignat}{2}
	 A_1 &= (1-n_x^2)^2~~,  & A_6 &= \frac{1}{2} \Bigl(4n_y^2-(3-n_x)(1-n_x)(1+n_x)^2\Bigr)~~, \nonumber \\
	 A_2 &= \frac{8n_z^2+n_y^2(1-n_x)^4}{(1-n_x)^2}~~, & A_7 &= \frac{1}{2} \Bigl(  (1-n_x)^3(1+n_x) -4n_y^2  \Bigr)~~, \nonumber \\
	 A_3 &= \frac{8n_z^2(3-n_x)+(1-n_x)^5}{16(1-n_x)}~~, &~~~ A_8 &= \frac{n_y}{2(1-n_x)^2} \Bigl(4n_z^2(3-n_x)+(1-n_x)^5\Bigr)~~, \\
	 A_4 &= \frac{8n_y^2(3-n_x)+(1-n_x)^5}{16(1-n_x)}~~, & A_9 &= \frac{n_y(n_x-3)}{2(1-n_x)^2} \Bigl(4n_z^2+(1-n_x)^3(1+n_x)\Bigr)~~, \nonumber\\
	 A_5 &= 2n_y (1-n_x)(1-n_x^2)~~, & A_{10} &= \frac{1}{8}(n_x-3)(1+n_x)(1-n_x)^2~~. \nonumber
\end{alignat}
To proceed we decompose 
\begin{equation}
	\tilde{\beta}_{\mu\nu} = a_{\mu\nu} + i ~b_{\mu\nu}~~,
\end{equation}
and find the following final expression:
\begin{equation}
	f_G(U,\Omega) = \frac{a^2 N(\Omega)^2}{4(U^2+a^2 \bar{\varepsilon}^2)^2} 
	\Bigl(
	 \alpha~  U^2 
	  + \beta~ U 
	  + \gamma
	\Bigr)~~,
\end{equation}
\begin{multline} \label{a1.2}
	\alpha = 
	A_1 a_{vv}^2
	+A_2 a_{vy}^2
	+A_3 a_{yy}^2
	+A_4 a_{zz}^2
	+A_5 a_{vv} a_{vy} +
	\\
	+A_6 a_{vv} a_{yy}
	+A_7 a_{vv} a_{zz}
	+A_8 a_{vy} a_{yy}
	+A_9 a_{vy} a_{zz}
	+A_{10} a_{yy} a_{zz}~~,
\end{multline}
\begin{multline}
	\beta = a \bar{\varepsilon} 
	\Bigl(
	2A_1 a_{vv} b_{vv}
	+2A_2 a_{vy} b_{vy} 
	+2A_3 a_{yy} b_{yy} 
	+2A_4 a_{zz} b_{zz} + \\
	+ A_5 (a_{vy} b_{vv} + a_{vv} b_{vy})  
	+ A_6 (a_{yy} b_{vv} + a_{vv} b_{yy}) 
	+ A_7 (a_{zz} b_{vv} + a_{vv} b_{zz})+ \\
	+ A_8 (a_{yy} b_{vy} + a_{vy} b_{yy})
	+ A_9 (a_{zz} b_{vy} + a_{vy} b_{zz})
	+ A_{10} (a_{zz} b_{yy} + a_{yy} b_{zz})\Bigr) ~~,
\end{multline} 
\begin{multline} \label{a1.4}
	\gamma = a^2 \bar{\varepsilon} ^2
	\Bigl(
	A_1 b_{vv}^2
	+A_2 b_{vy}^2
	+A_3 b_{yy}^2
	+A_4 b_{zz}^2
	+A_5 b_{vv} b_{vy}+ \\
	+A_6 b_{vv} b_{yy}
	+A_7 b_{vv} b_{zz}
	+A_8 b_{vy} b_{yy}
	+A_9 b_{vy} b_{zz}
	+A_{10} b_{yy} b_{zz} \Bigr)~~,
\end{multline}
where
\begin{equation}
	\begin{split}
		a_{vv} &= \frac{r_g}{2} (a_\omega - a_0 )~~, \\
		a_{vy} &= -r_g \omega a_\omega~~, \\
		a_{yy} &= r_g \left( - \frac{(1+n_x-2\omega n_y - \omega^2 (1-n_x)) }{1-n_x} a_\omega +  \frac{1+n_x}{1-n_x} a_0 \right) ~~,\\
		a_{zz} &= r_g \left( - \frac{(1+n_x-2\omega n_y + \omega^2 (1-n_x)) }{1-n_x}  a_\omega +  \frac{1+n_x}{1-n_x}  a_0 \right)~~,
	\end{split}
\end{equation}
\begin{equation}
	\begin{split}
		b_{vv} &= \frac{r_g}{2} (b_\omega - b_0 )~~, \\
		b_{vy} &= -r_g \omega b_\omega~~, \\
		b_{yy} &= r_g \left( - \frac{(1+n_x-2\omega n_y - \omega^2 (1-n_x)) }{1-n_x} b_\omega +  \frac{1+n_x}{1-n_x} b_0 \right)~~, \\
		b_{zz} &= r_g \left( - \frac{(1+n_x-2\omega n_y + \omega^2 (1-n_x)) }{1-n_x}  b_\omega +  \frac{1+n_x}{1-n_x}  b_0 \right)~~,
	\end{split}
\end{equation}
\begin{align}
	a_\omega = \Re~ \tilde{\phi}_\omega
	&= 
	\frac{ \sqrt{2(1-n_x)}\bigl( -n_y+\omega(1-n_x)\bigr)}{ \bar{\varepsilon} \Bigl(2(1-\omega n_y) + \omega^2(1-n_x)\Bigr)}~~, \\
	b_\omega = \Im~ \tilde{\phi}_\omega
	&=
	\frac{- \sqrt{2(1-n_x)}}{  \Bigl(2(1-\omega n_y) + \omega^2(1-n_x)\Bigr)} ~~,
\end{align}
and also $a_0 = a_{\omega=0}, b_0 = b_{\omega=0}$.

\section{Coordinate invariance of the averaged pseudo-tensor}\label{app2}
\setcounter{equation}0

Here we demonstrate that the averaged $t^r_U$-component of the pseudo-tensor (and, therefore, the averaged intensity of the GW flux)
is invariant under general coordinate transformations which preserve asymptotic \eqref{3.2} but may violate gauge conditions
\eqref{3.5}. Thus the presence of logarithmic terms does not affect this important property.

The coordinate transformations are generated by a vector field $\xi^\mu$. We suppose that large $r$ asymptotic
of the components of this vector in Minkowski coordinates is 
\begin{equation}\label{t5}
\xi^\mu=
 \alpha^\mu(\Omega)
+\tilde\alpha^\mu(\Omega)\ln( r/\varrho)
+r^{-1} \bigl(\sigma^\mu(U,\Omega)
+\tilde\sigma^\mu(U,\Omega) \ln ( r/\varrho) \bigr)
+O(r^{-2}\ln r)~~.
\end{equation}
Vectors $\alpha^\mu$, $\tilde{\alpha}^\mu$, $\sigma^\mu$, $\tilde{\sigma}^\mu$ are fairly arbitrary. We only require that
components $\sigma^\mu$, $\tilde\sigma^\mu$ are restricted at $U\to \pm \infty$.
After this transformation \eqref{3.2} can be written as follows:
\begin{equation}\label{t7}
h^{(1)'}_{\mu\nu}=r^{-1}
\Bigl(
H_{\mu\nu}^{(1)}(U,\Omega)+\Xi_{\mu\nu}(U,\Omega)
+\bigl(\tilde H_{\mu\nu}^{(1)}(\Omega)+\tilde\Xi_{\mu\nu}(U,\Omega)\bigr) \ln( r/\varrho) \Bigr)
+O(r^{-2}\ln r)~~,
\end{equation}
\begin{alignat}{2}
	\Xi_{UU}&=2\partial_U\sigma_U~~, & \tilde\Xi_{UU}&=2\partial_U\tilde\sigma_U~~, \nonumber \\
	\Xi_{Ur}&=\partial_U\sigma_r+\tilde\alpha_U~~, & \tilde\Xi_{Ur}&=\partial_U\tilde\sigma_r~~, \nonumber \\
	\Xi_{UA}&=\partial_A\alpha_U+\partial_U\sigma_A~~, & \tilde\Xi_{UA}&=\partial_A\tilde\alpha_U+\partial_U\tilde\sigma_A~~,  \label{t6} \\
	\Xi_{rr}&=2\tilde\alpha_r~~, & \tilde\Xi_{rr}&=0~~, \nonumber \\
	\Xi_{rA}&=\partial_A\alpha_r-\alpha_A+\tilde\alpha_A~~, & \tilde\Xi_{rA}&=\partial_A\tilde\alpha_r-\tilde\alpha_A~~, \nonumber\\
	\Xi_{AB}&=2\eth_{(A}\alpha_{B)}+2\gamma_{AB}(\alpha_r-\alpha_U)~~, &~~~~~~ \tilde\Xi_{AB}&=2\eth_{(A}\tilde\alpha_{B)}+2\gamma_{AB}(\tilde\alpha_r-\tilde\alpha_U)~~. \nonumber
\end{alignat}
A straightforward computation shows that $t^r_U$ component for \eqref{t7} is determined by the expression:
\begin{multline} \label{t8}
r^2(G^{(2)})_{U}^{r}(h^{(1)'})=N_{AB}N^{AB}-2\partial_U\left(N^{AB}\left[\frac12
C_{AB}+\eth_A\alpha_B+\gamma_{AB}(\alpha_r-\alpha_U)\right]\right)+
\\ +
2\partial_U\left(N^{AB}\left[\tilde
C_{AB}+\eth_A\tilde\alpha_B+\gamma_{AB}(\tilde\alpha_r-\tilde\alpha_U)\right]\right)\ln
( r/\varrho)+O(r^{-1}\ln^2 r)~~.
\end{multline}
Our averaging procedure implies integration over a sufficiently large time interval. As is easy to see from 
estimates \eqref{3.19a}, terms $\partial_U \left( N^{AB}[\ldots]\right)$ in the right hand side of \eqref{t8}
do not contribute at $U\to \pm \infty$.  Therefore the averaged $t^r_U$ and the flux do not depend on the chosen gauge condition.

If we restrict the class of coordinate transformations by requiring that they preserve the gauge conditions
\eqref{3.5}, which implies that
\begin{equation}\label{t2}
\Xi_{rr}=\tilde\Xi_{rr}=\Xi_{rA}=\tilde
\Xi_{rA}=\Xi_{AB}\gamma^{AB}=\tilde\Xi_{AB}\gamma^{AB}=0~~,
\end{equation}
it fixes arbitrary functions $\alpha_\mu$,
$\tilde\alpha_\mu$, as:
\begin{equation}\label{t9}
    \tilde
    \alpha_\mu=0~~,~~\alpha_A=\partial_A \alpha_r~~,~~\alpha_U=\left(\frac12\eth_A\eth^A+1\right)\alpha_r~~.
\end{equation}
Like in case of the BMS group the whole transformation is determined by a single function $\alpha_r(\Omega)=-\alpha^U(\Omega)$, related to the  supertranslation $U\to U-\alpha^U(\Omega)$.

\newpage
\bibliographystyle{unsrt}
%\bibliography{nullStr.bib}
%\end{document}

\end{document}